\documentstyle[mprocl,epsfig]{article}

\gdef\Feynmanlength{\setlength{\unitlength}{0.01pt}}  

\global\newcount\LINETYPE
\global\newcount\LINEDIRECTION
\global\newcount\LINECONFIGURATION
\newcommand{\LTYPE}{\LINETYPE}
\newcommand{\LDIR}{\LINEDIRECTION}

\global\LINETYPE=1  \global\LINEDIRECTION=0  \global\LINECONFIGURATION=0
\global\newcount\fermion    \fermion=1
\global\newcount\scalar     \scalar=2
\global\newcount\photon     \photon=3
\global\newcount\gluon      \gluon=4
\global\newcount\SPECIAL    \SPECIAL=5
\gdef\N{0}  \gdef\NE{1}  \gdef\E{2}   \gdef\SE{3}
       
\global\newcount\REG            \global\REG=0
\global\newcount\FLIPPED        \global\FLIPPED=1
\global\newcount\CURLY          \global\CURLY=2
\global\newcount\FLIPPEDCURLY   \global\FLIPPEDCURLY=3
\global\newcount\FLAT           \global\FLAT=4
\global\newcount\FLIPPEDFLAT    \global\FLIPPEDFLAT=5
\global\newcount\CENTRAL        \global\CENTRAL=6
\global\newcount\FLIPPEDCENTRAL \global\FLIPPEDCENTRAL=7
             
\global\newcount\SQUASHEDGLUON  \global\SQUASHEDGLUON=8

\newcount\adjx \adjx=0
\newcount\adjy \adjy=0
\global\newdimen\BIGPHOTONS     \BIGPHOTONS=0pt  
\global\newdimen\THICKPHOTONS     \THICKPHOTONS=0pt  
\global\newdimen\THICKPHOTONSWITCH    \THICKPHOTONSWITCH=0pt
\gdef\THICKPHOTONTEST{
\THICKPHOTONSWITCH=0pt
\ifdim\THICKPHOTONS=0pt \relax
  \else \ifnum\LTYPE=3
           \ifnum\LDIR=2 \THICKPHOTONSWITCH=1pt \fi 
           \ifnum\LDIR=6 \THICKPHOTONSWITCH=1pt \fi 
        \fi
\fi
}  
\gdef\THICKLINES{\thicklines  \THICKPHOTONS=1pt}

\global\newcount\phantomswitch   \global\phantomswitch=0
\global\newcount\stemlength   \global\stemlength=275   
\global\newcount\absstemlength        
\global\newcount\stemlengthx          
\global\newcount\stemlengthy          
\newdimen\FRONTSTEM  \FRONTSTEM=0pt   
\newdimen\BACKSTEM   \BACKSTEM=0pt    
\newdimen\EITHERSTEM \EITHERSTEM=0pt  
\global\newcount\arrowlength                
\global\newdimen\ATTIP   \global\ATTIP=0pt  
\global\newdimen\ATBASE  \global\ATBASE=1pt 
\global\newcount\unitboxnumber  
\global\newcount\unitboxnumberpo  
\global\newcount\particlelengthx  
\gdef\plengthx{\particlelengthx}
\global\newcount\particlelengthy  
\gdef\plengthy{\particlelengthy}
\global\newcount\boxlengthx  
\global\newcount\boxlengthy  
\global\newcount\particleadjustx  
\global\newcount\particleadjusty  
\global\newcount\particlelength   
\global\newcount\particlefrontx
\gdef\pfrontx{\particlefrontx}
\global\newcount\PFRONTx
\global\newcount\particlefronty
\gdef\pfronty{\particlefronty}
\global\newcount\PFRONTy
\global\newcount\particlebackx
\gdef\pbackx{\particlebackx}
\global\newcount\particlebacky
\gdef\pbacky{\particlebacky}
\global\newcount\particlemidx
\gdef\pmidx{\particlemidx}
\global\newcount\particlemidy
\gdef\pmidy{\particlemidy}
\global\newcount\seglength  \global\newcount\gaplength
\global\gaplength=850  
\global\seglength=1416  
\global\newcount\Xone    \global\newcount\Yone    
\global\newcount\Xtwo    \global\newcount\Ytwo    
\global\newcount\Xthree  \global\newcount\Ythree  
\global\newcount\Xfour   \global\newcount\Yfour   
\global\newcount\Xfive   \global\newcount\Yfive   
\global\newcount\Xsix    \global\newcount\Ysix    
\global\newcount\Xseven  \global\newcount\Yseven  
\global\newcount\Xeight  \global\newcount\Yeight  
\newsavebox{\lastline}  
\global\newcount\numlineparts   
\global\newcount\upperlineadjx  \upperlineadjx=0  
\global\newcount\upperlineadjy  \upperlineadjy=0  
\global\newcount\lowerlineadjx  \lowerlineadjx=0  
\global\newcount\lowerlineadjy  \lowerlineadjy=0  
\global\newcount\thirdlineadjx  \thirdlineadjx=0  
\global\newcount\thirdlineadjy  \thirdlineadjy=0  
\global\newcount\fourthlineadjx \fourthlineadjx=0  
\global\newcount\fourthlineadjy \fourthlineadjy=0  
\global\newcount\unitboxwidth   \unitboxwidth=1000
\global\newcount\unitboxheight  \unitboxheight=0  
\global\newcount\numupperunits  \numupperunits=8  
\global\newcount\numlowerunits  \numlowerunits=8  
\global\newcount\numthirdunits  \numthirdunits=8  
\global\newcount\numfourthunits \numfourthunits=8  
\global\newcount\fermioncount   \global\fermioncount=0
\global\newcount\scalarcount    \global\scalarcount=0
\global\newcount\photoncount    \global\photoncount=0
\global\newcount\gluoncount     \global\gluoncount=0
\global\newcount\SPECIALcount   \global\SPECIALcount=0
\global\newcount\vertexcount    \global\vertexcount=-1
\global\newcount\XDIR
\global\newcount\YDIR
\gdef\SETDIR{  
\ifcase\LDIR
     \global\XDIR=0  \global\YDIR=1   
\or  \global\XDIR=1  \global\YDIR=1   
\or  \global\XDIR=1  \global\YDIR=0   
\or  \global\XDIR=1  \global\YDIR=-1  
\or  \global\XDIR=0  \global\YDIR=-1  
\or  \global\XDIR=-1 \global\YDIR=-1  
\or  \global\XDIR=-1 \global\YDIR=0   
\or  \global\XDIR=-1 \global\YDIR=1   
\else\DIRECTERROR
\fi}  
\gdef\moduloeight#1{
\ifnum#1>7 \global\advance #1 by -8
\relax
\moduloeight#1
\relax
\else \relax
\fi}
\gdef\multroothalf#1{\global\multiply #1 by 7071 \global\divide #1 by 10000}
\gdef\negate#1{\global\multiply #1 by -1}

\gdef\slanttest(#1,#2){
\ifodd\LDIR
\multiply #1 by 7071  \divide #1 by 10000
\multiply #2 by 7071  \divide #2 by 10000
\fi
}
\gdef\gslanttest(#1,#2){
\ifodd\LDIR
\multroothalf#1
\multroothalf#2
\fi
}
\gdef\setplength{ 
\global\particlelengthx=\unitboxwidth
\global\particlelengthy=\unitboxheight
\global\multiply \particlelengthx by \unitboxnumber
\global\multiply \particlelengthy by \unitboxnumber
\global\advance \particlelengthx by \particleadjustx
\global\advance \particlelengthy by \particleadjusty
}
\gdef\boxlengthdefault{  
\global\boxlengthx=\plengthx
\global\boxlengthy=\plengthy
\ifnum\plengthx<0 \global\multiply\boxlengthx by -1 \fi
\ifnum\plengthy<0 \global\multiply\boxlengthy by -1 \fi
}
\gdef\rearcoords{  
\global\particlebacky=\particlefronty
\global\particlebackx=\particlefrontx
\global\advance \particlebackx by \particlelengthx
\global\advance \particlebacky by \particlelengthy
}
\gdef\midcoords{  
\global\particlemidy=\particlefronty
\global\particlemidx=\particlefrontx
\global\stemlengthx=\particlelengthx  
\global\stemlengthy=\particlelengthy
\global\divide\stemlengthx by 2
\global\divide\stemlengthy by 2
\global\advance \particlemidx by \stemlengthx
\global\advance \particlemidy by \stemlengthy
}
\gdef\setparticle{\setplength\rearcoords\midcoords\boxlengthdefault}  
\gdef\setcoords(#1,#2,#3)(#4,#5,#6)[#7,#8]{
\global\upperlineadjx=#1
\global\lowerlineadjx=#2
\global\thirdlineadjx=#3
\global\upperlineadjy=#4
\global\lowerlineadjy=#5
\global\thirdlineadjy=#6
\global\unitboxwidth=#7
\global\unitboxheight=#8
}
\gdef\drawoldpic#1(#2,#3){  
\global\particlefrontx=#2
\global\particlefronty=#3
\rearcoords
\midcoords
\put(#2,#3){\usebox{#1}}
}
\gdef\drawsavedline`#1' as #2[#3#4](#5,#6)[#7]{
\global\LINETYPE=#2
\global\LINEDIRECTION=#3
\global\LINECONFIGURATION=#4
\global\particlefrontx=#5
\global\particlefronty=#6
\global\unitboxnumber=#7
\selectcase
\rearcoords
\midcoords
\ifnum\phantomswitch=0 \drawas{#1}\fi
}

\gdef\drawas#1{
\global\savebox{#1}(\boxlengthx,\boxlengthy){
\setlength{\unitlength}{0.01pt}
\begin{picture}(\boxlengthx,\boxlengthy)
\multiput(\upperlineadjx,\upperlineadjy)(\unitboxwidth,\unitboxheight)
{\numupperunits}{\upperunitbox}
\ifnum\numlineparts > 1  
\multiput(\lowerlineadjx,\lowerlineadjy)(\unitboxwidth,\unitboxheight)
{\numlowerunits}{\lowerunitbox}
\fi
\ifnum\numlineparts > 2  
\multiput(\thirdlineadjx,\thirdlineadjy)(\unitboxwidth,\unitboxheight)
{\numthirdunits}{\thirdunitbox}
\fi
\ifnum\numlineparts > 3  
\multiput(\fourthlineadjx,\fourthlineadjy)(\unitboxwidth,\unitboxheight)
{\numfourthunits}{\lowerunitbox}
\fi
\end{picture} }
\global\PFRONTx=\pfrontx  \global\PFRONTy=\pfronty   
\SETFRONTSTEM
\THICKPHOTONTEST
\ifdim\THICKPHOTONSWITCH=1pt\global\advance\PFRONTy by 20  \fi
\put(\PFRONTx,\PFRONTy) {\usebox{#1}}   
\ifdim\THICKPHOTONSWITCH=1pt
\global\advance\PFRONTy by -40
\put(\PFRONTx,\PFRONTy) {\usebox{#1}}   
\global\advance \PFRONTy by 20  
\fi  
\SETBACKSTEM
\seglength=1416   \gaplength=850   
}
\gdef\drawandsaveline`#1' as #2[#3#4](#5,#6)[#7]{
\global\newsavebox{#1}
\drawsavedline`#1' as #2[#3#4](#5,#6)[#7]
}

\gdef\drawline#1[#2#3](#4,#5)[#6]{   
\drawsavedline`\lastline' as #1[#2#3](#4,#5)[#6]}

\gdef\TYPEERROR{\message{*** ERROR IN PARTICLE TYPE SELECTION ***}
\message{+++ Try with line type \fermion,\scalar,\photon,\gluon
(see manual) +++}\SETERR}
\gdef\DIRECTERROR{\SETERR\message{*** ERROR IN PARTICLE DIRECTION SELECTION
***}
\message{+++ Try again with direction N, NE, E, SE  etc. or see manual +++}}
\gdef\UNIMPERROR{\message{*** ERROR IN PARTICLE OPTIONS SELECTION ***}
\message{
+++ The requested options combination has not yet been implemented +++}\SETERR}
\gdef\SETERR{\gdef\upperunitbox{{\tiny Error}}  
\gdef\lowerunitbox{\relax}
\gdef\thirdunitbox{\relax}
}
\gdef\neglengthcheck{\ifnum\unitboxnumber < 1
\message{   *** ERROR:  PARTICLE OF NEGATIVE OR ZERO LENGTH REQUESTED. ***   }
\message{   ***         TAKING ABSOLUTE VALUE. ***   }\negate\unitboxnumber
\fi}
\gdef\selectcase{
\neglengthcheck   
\SETDIR
\ifcase\LINETYPE
\TYPEERROR  
\or \selectfermion  
\or \selectscalar   
\or \selectphoton   
\or \selectgluon    
\or \selectspecial  
\else \TYPEERROR \fi  }
\gdef\selectfermion{
\ifnum\fermioncount=0 
\global\newcount\fermionlength  
\global\newcount\fermionlengthx
\global\newcount\fermionlengthy
\global\newcount\fermionfrontx  
\global\newcount\fermionfronty  
\global\newcount\fermionbackx
\global\newcount\fermionbacky
\gdef\ALLfermion{  
\global\fermionfrontx=\particlefrontx \global\fermionfronty=\particlefronty
\ifnum\unitboxnumber > 50000
\message{   *** WARNING *** Fermion of length
\the\unitboxnumber\space requested ***   }
\ifnum\unitboxnumber > 80000
\message{   *** Reducing fermion length to 30000 (max 80000) ***   }
\global\unitboxnumber=30000 \fi \fi  
\global\fermionlength=\unitboxnumber 
\global\particleadjustx=0   \global\particleadjusty=0 
\global\numlineparts = 1    \global\numupperunits=1
\global\upperlineadjx=-200  \global\upperlineadjy=0
\global\fermionlengthx=\fermionlength    \global\fermionlengthy=\fermionlength
\gslanttest(\fermionlengthx,\fermionlengthy)  
\global\multiply\fermionlengthx by \XDIR  
\global\multiply\fermionlengthy by \YDIR  
\global\unitboxheight=\fermionlengthy   \global\unitboxwidth=\fermionlengthx
\global\advance \fermionlengthx by \particleadjustx
\global\advance \fermionlengthy by \particleadjusty
\global\particlelengthx=\fermionlengthx
\global\particlelengthy=\fermionlengthy
\boxlengthdefault    \rearcoords    \midcoords
\global\fermionbackx=\particlebackx     \global\fermionbacky=\particlebacky
\ifcase\LINECONFIGURATION  
\ifnum\XDIR=0
\gdef\upperunitbox{\line(\XDIR,\YDIR){\boxlengthy}} 
\else
\gdef\upperunitbox{\line(\XDIR,\YDIR){\boxlengthx}}
\fi
\else \UNIMPERROR
\fi
}

 \fi
\global\advance\fermioncount by 1  
\ALLfermion
}
\gdef\selectscalar{
\ifnum\scalarcount=0 
\newcount\scalarlength
\newcount\scalarlengthx
\newcount\scalarlengthy
\newcount\scalarfrontx  
\newcount\scalarfronty  
\newcount\scalarbackx
\newcount\scalarbacky
\gdef\ALLscalar{
\global\scalarfrontx=\particlefrontx   
\global\scalarfronty=\particlefronty   
\numlineparts = 1      \numupperunits=\unitboxnumber
\ifcase\LINECONFIGURATION
\global\upperlineadjx=-200     \global\upperlineadjy=0
\slanttest(\seglength,\gaplength)   
\gdef\upperunitbox{\line(\XDIR,\YDIR){\seglength}}
\else \UNIMPERROR 
\fi
\global\unitboxwidth=\seglength  \global\advance\unitboxwidth by \gaplength
\global\multiply \unitboxwidth by \XDIR
\global\unitboxheight=\seglength  \global\advance\unitboxheight by \gaplength
\global\multiply \unitboxheight by \YDIR
\global\particleadjustx=\gaplength \global\multiply\particleadjustx by \XDIR
\global\particleadjusty=\gaplength \global\multiply\particleadjusty by \YDIR
\negate\particleadjustx   \negate\particleadjusty   
\setparticle  
\global\scalarlengthx=\particlelengthx  
\global\scalarlengthy=\particlelengthy  
\ifnum\boxlengthx > 50000
\message{   *** WARNING *** Scalar of length in excess of 50000cp
requested!}\fi
\ifnum\boxlengthy > 50000
\message{   *** WARNING *** Scalar of length in excess of 50000cp
requested!}\fi
\global\scalarbackx=\pbackx      \global\scalarbacky=\pbacky   
}

 \fi
\global\advance\scalarcount by 1  
\ALLscalar
}
\gdef\selectphoton{   
\ifnum\photoncount=0 \input PHOTONSETUP  \fi
\selectphoton
}
\gdef\selectgluon{   
\ifnum\gluoncount=0 \input GLUONSETUP  \fi
\selectgluon
}
\gdef\selectspecial{\UNIMPERROR}
\gdef\checkvertex{ 
\ifnum\vertexcount=-1   \input VERTEX  \fi}
\gdef\drawvertex#1[#2#3](#4,#5)[#6]{\checkvertex\drawvertex#1[#2#3](#4,#5)[#6]}
\gdef\vertexcap#1{\checkvertex\vertexcap#1}
\gdef\vertexcaps{\checkvertex\vertexcaps}
\gdef\vertexlink#1{\checkvertex\vertexlink#1}
\gdef\vertexlinks{\checkvertex\vertexlinks}
\gdef\stemvertex#1{\checkvertex\stemvertex#1}
\gdef\stemvertices{\checkvertex\stemvertices}
\gdef\flipvertex{\checkvertex\flipvertex}
\global\arrowlength=349  
\gdef\drawarrow[#1#2](#3,#4){
\global\LDIR=#1
\SETDIR
\global\boxlengthx=#3  
\global\boxlengthy=#4  
\ifdim#2=1pt  
\adjx=\arrowlength      \adjy=\arrowlength
\multiply\adjx by \XDIR \multiply\adjy by \YDIR  
\slanttest(\adjx,\adjy)
\global\advance\boxlengthx by \adjx    \global\advance\boxlengthy by \adjy
\fi
\ifnum\phantomswitch=0\put(\boxlengthx,\boxlengthy){\vector(\XDIR,\YDIR){0}}\fi
}  
\gdef\SETFRONTSTEM{
\EITHERSTEM=\FRONTSTEM   \advance\EITHERSTEM by \BACKSTEM
\ifdim\EITHERSTEM>0pt
\global\stemlengthx=\stemlength   \global\stemlengthy=\stemlength
\global\absstemlength=\stemlength
\SETDIR
\gslanttest(\stemlengthx,\stemlengthy)
\gslanttest(\absstemlength,\REG)  
\ifnum\XDIR=0 \stemlengthx=0 \fi
\ifnum\YDIR=0 \stemlengthy=0 \fi
\global\multiply\stemlengthx by \XDIR
\global\multiply\stemlengthy by \YDIR
\ifdim\FRONTSTEM=1pt
\ifnum\phantomswitch=0
          \put(\pfrontx,\pfronty){\line(\XDIR,\YDIR){\absstemlength}}\fi
\global\advance\plengthx by \stemlengthx
\global\advance\plengthy by \stemlengthy
\global\advance\PFRONTx by \stemlengthx
\global\advance\PFRONTy by \stemlengthy
\global\advance\pmidx by \stemlengthx
\global\advance\pmidy by \stemlengthy
\global\advance\pbackx by \stemlengthx
\global\advance\pbacky by \stemlengthy
\ifnum\LTYPE=3
\global\photonfrontx=\PFRONTx  \global\photonfronty=\PFRONTy
\global\photonbackx=\pbackx    \global\photonbacky=\pbacky
\fi  
\ifnum\LTYPE=4
\global\gluonfrontx=\PFRONTx  \global\gluonfronty=\PFRONTy
\global\gluonbackx=\pbackx    \global\gluonbacky=\pbacky
\fi  
\fi  
\fi  
}    
\gdef\SETBACKSTEM{
\ifdim\BACKSTEM=1pt
\ifnum\phantomswitch=0
       \put(\pbackx,\pbacky){\line(\XDIR,\YDIR){\absstemlength}}\fi
\global\advance\plengthx by \stemlengthx
\global\advance\plengthy by \stemlengthy
\global\advance\pbackx by \stemlengthx
\global\advance\pbacky by \stemlengthy
\fi  
\global\stemlength=275  \FRONTSTEM=0pt  \BACKSTEM=0pt 
}    
\gdef\drawloop#1[#2#3](#4,#5){  
\input LOOPS  
\drawloop#1[#2#3](#4,#5)}
\Feynmanlength  

\def\Journal#1#2#3#4{{#1} {\bf #2}, #3 (#4)}

\def\NPB{{\em Nucl. Phys.} B}
\def\PLB{{\em Phys. Lett.}  B}
\def\PRL{\em Phys. Rev. Lett.}
\def\PRD{{\em Phys. Rev.} D}

\def\be{\begin{equation}}
\def\ee{\end{equation}}
\def\bea{\begin{eqnarray}}
\def\eea{\end{eqnarray}}

\begin{document}
\title{ STANDARD MODEL PARAMETERS FROM THE MULTIPLE POINT
PRINCIPLE AND ANTI-GUT
}

\author{ D.L. BENNETT, H.B. NIELSEN }

\address{ The Niels Bohr Institute,
Copenhagen {\O}, Denmark}

\author{  C.D. FROGGATT }

\address{ Department of Physics and Astronomy,
 Glasgow University, Glasgow G12 8QQ, Scotland}


\maketitle\abstracts{We put forward a model, or rather
a relatively broad class of
models, beyond the Standard Model based on the
two main assumptions: MPP) The coupling constants should be fixed
such as to ensure that there be
many ``vacuum states'', i.e.~Lorentz invariant
states of the fields, with the same energy density.
Anti-GUT) At high energy, above an essential desert of only
Standard Model interactions,
there is the bigger gauge group $SMG^3\times U(1)_f$ which means that
each family of quarks and leptons has its own set of gauge bosons
analogous to those in the Standard Model
itself, and then there is one extra
abelian gauge boson called $U(1)_f$. In
addition we make some further more
phenomenological assumptions. We succeed in fitting order of magnitudewise
most of the (effective) Yukawa couplings observed in the
Standard Model as quark and lepton masses
and mixing angles. We
have more accurate numbers for the three fine-structure constants
and the top and Higgs masses, as well as some suggestive understanding
of the fine-tuning wonders for the cosmological constant
and the $\Theta$-angles.
CP-violation is predicted a bit low, but order of magnitudewise
in agreement with experiment. In summary,
we obtain
at least an order of magnitudewise understanding of
the Standard Model parameters within our scheme,
except for the mysteriously low weak scale compared to the Planck scale,
with only 4 continuous parameters being fitted,
in addition to our somewhat more discrete choices.}

\section{Introduction}\label{sec:intro}

If one wishes to seek inspiration from experimental data for
clues to the theory beyond the Standard Model,
one has not much else to work with other than
the about 19 Standard Model parameters not fixed by the
Standard Model itself.
Otherwise, one only has information from
neutrino oscillation experiments
and cosmological studies, together with the
lack of evidence for baryon decay or
other new physics deviating from the Standard Model.
In the present talk we would like to specify,
at least order of magnitudewise,
these parameters in a model for the physics beyond the Standard Model.
We do not really use a full model, but rather a series of assumptions
which partially fix the model \cite{picek,book,glasgowbrioni,db3,db2}.
The two most characteristic assumptions---we
discuss our assumptions in more detail below---for
this ``fitting'' of the Standard Model parameters are:

\begin{description}
\item[1. MPP.] The parameters (coupling constants)
must have their values adjusted so that a large number of vacuum states
have the same energy density.
In fact it is easy to show that the couplings, for which two
vacua are degenerate, are just the same ones
for which the Euclideanised
(imaginary time) version of the theory has a phase transition.
So if several vacua are degenerate there is a multiple point.
Hence we call our assumption the
Multiple Point Principle: MPP.

\item[2. Anti-GUT.] Near the Planck scale, the gauge group is
the anti-grand unified
group $ SMG^3\times U(1)_f$. This means that each of the three
quark-lepton generations
have got their own gauge group with
the same
structure as the Standard Model
Group $SMG$. In
addition there is an abelian flavour group, which we call $U(1)_f$.

\end{description}

In outline our model runs roughly like this:
The pure Standard Model, together with its yet to be found
Weinberg-Salam Higgs particle,
will be valid with high accuracy up to higher energies
than most other physicists
believe; namely up to about one order of magnitude below the Planck
energy scale. That is to say our model
would be falsified by the discovery of,
for example, supersymmetric partners in any
experimentally accessible mass range.
Actually the existence of such partners would disturb
some of the agreements of our model with experiment.
Then up about one order
of magnitude below the Planck scale,
one should find, in our picture, that the Standard Model Group,
called for short $SMG$, is the ``diagonal'' subgroup of a
much bigger gauge
group---the Anti-GUT postulate.
This non-simple anti-grand
unified group $ SMG^3\times U(1)_f$
is supposed to be broken down by a series
of Higgs fields, to which we have given the names $W$, $T$, $\xi$ and $S$.

The Grand Unified SU(5) group, often thought to exist
beyond the Standard Model,
is not a subgroup of our group $SMG^3\times U(1)_f$. So we do not have
SU(5) in our model
and the coincidence of the gauge couplings agreeing
with the SU(5) prediction, after
supersymmetry (SUSY) corrections,
must be declared a total
accident or explained in a roundabout way, if at all,
in our model. Ironically
enough, our central predictions for the gauge couplings happen
to agree better
with simple SU(5) than the Standard Model experimental couplings do.
However that
is a kind of pure accident, since our calculational accuracy is too low to
spot deviations of the order of the SUSY corrections to the
gauge coupling differences at the SU(5) scale.
This accidental coincidence with the SU(5)-GUT result
is seen in figure 1, which shows
our anti-GUT predictions at the Planck
scale $M_{Planck} = 1.2 \times 10^{19}$ GeV.
Note that these
predictions are for the absolute values of the gauge couplings and not
just their ratios.

Above the anti-GUT breaking energy scale
(an order of magnitude below the Planck scale),
there are really three times as many gauge couplings.
For example the colour SU(3) group is replaced, at these scales,
by three SU(3) groups meaning 3 times 8 gluons,
one set of 8 gluons coupling only to the first generation,
the next set of 8 gluons only to the second generation,
and so on. In addition there is a quite extra U(1) gauge group, which
we refer to as $U(1)_f$.
In summary we can say: each generation has got
its own set of gauge fields/particles quite analogous to those in the
Standard Model, i.e. twelve gauge fields for each generation, only
coupling to just that generation, and then in addition
the $U(1)_f$ gauge field.
Altogether there are thus 3 x 8 = 24 gluons, 3 x 3 = 9 W's
and 3 x 1 + 1 = 4 abelian gauge bosons.

Contrary to many popular unifying
gauge group proposals, our group is very far from being in the
group theoretical sense a simple group; it has lots of nontrivial
invariant subgroups.  But, in our model, we do not need any
unification of the group to make the number of independent gauge
coupling constants small. We have a completely different method
of predicting the values of coupling constants. We postulate---and
this is the MPP postulate---that the various
coupling constants, for some reason or another,
have put themselves to just such values that a
lot of different Lorentz invariant states of the space
and its fields have got the same energy density and pressure.
They are degenerate.
Naturally the imposition of such a number of equalities among the
pressures, or equivalently of the
energy densities, of these vacuum states
leads to restrictions between the various parameters
(coupling constants) of the
field theory model. It is these restrictions that replace for us
the GUT predictions relating coupling constants
due to unification into simple groups.
Our predictions come instead from
imposing the common energy densities
for the supposed many vacua.

\begin{figure}
\leavevmode
\centerline{\epsfig{file=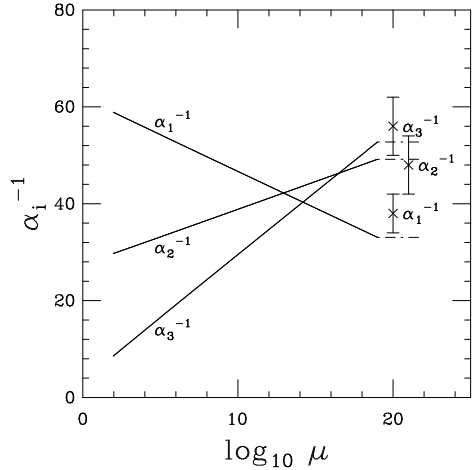,width=6.5cm}
}
\caption{Evolution of the Standard Model
fine structure constants $\alpha_i$ ($\alpha_1$
in the SU(5)
inspired normalisation)
from the electroweak scale to the Planck scale. The anti-GUT model
predictions for the values at the Planck scale,
$\alpha_i^{-1}(M_{Planck})$,
are shown with error bars.}
\label{fig:alphas}
\end{figure}

It is important for some of these predictions,
and it is a part of our model,
that there is a fundamental length of the order of magnitude of the Planck
length. At that scale there is,
what we would like to think of as, a truly
existing regularization. We sometimes think
and calculate as if it were due to
a truly physically existing lattice.
However we hope and speculate that, for most purposes,
it does not matter what the regularization
really is: a lattice, superstring theory, or yet another type
of cut-off. But it is important that we have
one or another cut-off, in as far as
some of the vacua declared degenerate are lattice (or regularization)
artifacts. So the idea is in our model, in which some sort of
regularization should really exist, that
such artifacts should also really exist.
Furthermore we shall require, for some as yet not
fully understood physical reason, the various regularization
artifact vacua to be degenerate; the corresponding
requirements on the coupling constants
are then expected to be valid in nature
(and should be looked for experimentally).

We shall even make an assumption of the type that up near this
regularization scale (taken to be the Planck scale), we can
essentially find particles with any
quantum numbers; so that all allowed amplitudes will exist
and correspond to rates
of order unity in Planck units.

It should be stressed that our model, especially at the high energy scale,
is not so terribly specific, but rather in some sense encompasses a large
class of models. We speculate, as an extra assumption,
that all the models in this class in practice behave in
much the same way. We really do not want to commit ourselves
to a specific regularization---lattice or superstring
or momentum cut-off say---and we do not really exclude
the possible existence of further gauge fields
not interfering with the $SMG^3\times U(1)_f$ gauge fields.
We do not try to specify precisely
the large number of possible particles
that should be found
at the Planck scale.
Only the somewhat lighter
than Planck mass particles are restricted strongly by our assumptions.

The fundamentally existing regularization could perhaps
just be that, at Planck scale distances,
the theory is really some superstring
theory which is free of divergencies. But to have agreement of
our predictions with the experimental couplings,
the top quark mass etc.,
we cannot tolerate supersymmetry to remain unbroken down
to even near the electroweak scale.
SUSY must be broken already close to the Planck scale.
Also our
group $SMG^3\times U(1)_f$
has to be included
in the gauge group of such a string model.
Moreover the string gauge group must break down to just
our anti-GUT group at the Planck scale,
except perhaps for some fully decoupled
extra groups.

Let us now spell out some of the details of the above sketched
picture of our model, by describing its
five most important features or assumptions:

\hspace{-0.5 cm}{\bf MPP: Many degenerate vacua.}

In a quantum field theory
there can a priori be several Lorentz
invariant states. The typical
example is a theory with a scalar
field for which the effective potential
has more than one minimum. For each
minimum there is a Lorentz invariant
vacuum state in which the scalar field
expectation value lies in the
chosen minimum. A priori these different
vacua have energy momentum tensors
proportional to the metric tensor,
but with different coefficients
for the different vacua. They have
different cosmological constants, one
could say. These different
cosmological constants can formally be
calculated for each proposed vacuum
and will turn out normally to
depend on the various coupling
constants (parameters) of the field
theory in question. If it were
not because of our assumption of a
physically existing regularization,
a divergent nonsensical number would usually result.
However with regularization
one should get a meaningful number.

Our assumption of degenerate vacua
now says that the coupling constants
take just such values that the
various cosmological constants for the
different vacua become equally large. That obviously should give
some equations among the various
coupling constants, depending of course
on the details of the field theory used.

But why should we expect that
these vacua should have the same energy
densities? Perhaps the most convincing
argument is provided by the successful phenomenological
predictions for the gauge
coupling constants obtained by using this requirement.
In addition we can list four other arguments which carry
less weight, but anyway helped motivate us
and may eventually lead to a deeper understanding
of our approach.

1) An argument essentially due to Susskind:
If there is some reason why the
cosmological constant in Nature (our vacuum)
is so accurately zero, then
the same argument is expected to work
for the other potential vacua. However
it should be admitted, for example, that if
the mechanism of Tsamis and Woodard
\cite {tsamis} making the cosmological zero is the reason,
Woodard denies that it should
also work to make the cosmological constant zero
for the scalar field in another minimum. So it
seems that the argument does
not work for all imaginable mechanisms
explaining the vanishing of the cosmological constant.

2) Ice-water analogy: In a
microcanonical ensemble for
water, i.e. with a definite amount of energy,
it does not take so terribly
much luck to get such an energy chosen
that the ensemble cannot be realized
(under the given pressure)
as pure liquid water, or as pure ice.
It has to be a mixture---slush---but then the temperature
has to have a fine-tuned
value equal to zero degrees Celsius, or rather the
freezing temperature. So
actually it can very easily happen that
slush is produced and that thus
the freezing temperature is very likely to occur.
This is due to the large latent heat of fusion of water giving a
strongly first order phase transition between ice and water.
By analogy with this
we want to say that one
expects it would be
very likely---using
somehow similar assumptions for the field theory---that the
microcanonical ensemble-like mechanism
would work and lead to precisely those couplings
for which the different vacua become
degenerate. The degeneracy of the vacua
is supposedly the analogous fine-tuning to slush being
at the freezing point of water.

3) Non-locality: If we generalize the
mixed phase (slush) mechanism, as
just suggested above, from the case of the statistical
mechanics partition function in three dimensions to the
four dimensional functional integral
formulation of quantum field theory,
it turns out that the analogue of the microcanonical ensemble
is strictly speaking non-local.
According to the principle of locality, the action $S$ used
to define the Feynman path integral
\begin{equation}
\int {\cal D}A {\cal D}\psi {\cal D} \phi \exp(iS[A,\psi,\phi])
\end{equation}
should be an integral over four dimensional space-time
of a Lagrangian density.
However, the analogue of the microcanonical ensemble
is obtained by replacing the exponentiated action $\exp(iS)$,
in the integrand of the path integral, by
a delta function of an integral over all space-time:
\begin{equation}
\int {\cal D}A {\cal D}\psi {\cal D} \phi \delta (I[A,\psi,\phi]-I_0)
\label{mipathway}
\end{equation}
where
\begin{equation}
I[A,\psi,\phi] = \int d^4x {\cal L}(x)
\label{expr3}
\end{equation}
But such a delta
function does not obey the principle of locality
and, actually, it makes the presence of degenerate vacua
a likely occurrence \cite{smtop,db2}.
If you rewrite the delta function path
integral eq. (\ref{mipathway})
as one with an exponentiated action
as usual, you would have to make
the action a non-linear function of
the expression in eq. (\ref{expr3}).
Alone letting the action be a non-linear
function of space-time integrals
like eq. (\ref{expr3}) is sufficient to
make it likely that there will be
degenerate vacua \cite{glasgowbrioni}, i.e. MPP.
Non-linearity in such expressions means non-locality.
With this form of mild non-locality the theory can for practical
purposes be simulated by a local theory \cite{book}.
The only left-over signal that
it is fundamentally non-local is the MPP, that is to say
the effective coupling constants
of the simulating local theory will
turn out to be adjusted to make several degenerate vacua.

4) Supersymmetry (but strongly broken): Supersymmetry
models typically have
several states of vacuum type with precisely the
same energy density, namely zero. In this way
the situation required by our MPP principle naturally
arises: the existence of several degenerate vacua.
However we must, as must also be done
for purely phenomenological reasons, assume that
supersymmetry is broken.
As we shall see below in section 6,
the supersymmetry breaking
scale must be quite close to the Planck scale,
so as to avoid spoiling the otherwise
good agreement between
the experimental gauge coupling constants and our model.

\hspace{-0.5 cm}{\bf Anti-GUT gauge group.}

Before defining our Anti-GUT {\em group},
it should be mentioned that we follow
Michel and O'Raifeartaigh \cite{Michel} in making sense not
only of the Lie algebra
but even of the Lie {\em group} for a
Yang-Mills theory. The interpretation
of the global properties of the gauge
group is that the matter field
representations should be only
those that could be representations of the
gauge {\em group}. As a familiar example you may think about
the rotation group; while the group
SU(2) has both half integer and
integer spin representations, the
group SO(3), having the same Lie
algebra as SU(2), has only the integer
spins as its group representations.
So we could  pack the information
that some set of particles has only integer spin
into the statement that they
are faithful representations of SO(3), but
not of SU(2). In an analogous way,
O'Raifeartaigh points out that only
the representations of the Standard Model
Lie algebra which also obey the charge quantisation rule
\begin{equation}
y/2 + d/2 + t/3 = 0 \quad (\mbox{mod} \quad 1)
\label{chargeqr}
\end{equation}
are representations of the
group $SMG$ = $S(U(3)\times U(2))$, which has the
same Lie algebra as $ SU(3) \times SU(2)\times U(1)$.
Here we used the definition of triality $t$ as the
quantum number, counted modulo 3, with the property
that it is related to the representation
of the particle type in question
under the colour group SU(3): if
the SU(3) representation is the triplet
representation the triality is
defined to be 1 (mod 3), and if it occurs
in the representation obtained by putting together $t$
triplet representation particles we say it has triality $t$.
Analogously the ``duality'' $d$ is defined
as a function, modulo 2, of the
representation under the SU(2) algebra:
if the weak isospin is half integer $d$
is odd (say $d=1$), and if the weak isospin is integer
we define $d$ to be even, $d=0$. The weak hypercharge $y/2$
is normalised
so that, for example, $y/2 = 1$ for the positron.
According to the well-known relation $Q= y/2 + T_3$
for the electric charge $Q$, where $T_3$ is the third
component of the weak isospin, the charge quantisation
rule eq. (\ref{chargeqr}) expresses
the somewhat complicated way in which the electric charge
is quantised in the Standard Model.

One may think of such quantisation rules
as the restriction of the
Lie algebra representations allowed by the introduction
of some type of monopole.
There is a direct correspondence between the
charge of the magnetic monopole and the
invariant discrete subgroup elements
to be ``divided out'' of the covering group, in order to
get to the {\em group} in question with the prescribed Lie algebra.

It should also be kept in mind that, in line with our philosophy
of a real physically existing
cut-off, e.g. a lattice (see below),
actually the gauge {\em group} and not only the Lie algebra acquires a
physical significance.

We therefore think of the extension of the Standard Model into
our anti-GUT model as an extension of
the {\em group} and not only of the Lie algebra.
So we also propose an extension of the charge quantisation
rule, at the same time as we extend the algebra.

Our proposed group is literally taken as
a {\em group} $SMG^3\times U(1)$ (and not
only an algebra). That is to say that for each
of the three $SMG$=$S(U(2)\times U(3))$ factors in the
cross product---each supposed to couple to just one generation,
leaving the other two untouched---there is a separate
quantisation rule just for that generation.
In addition the extra U(1)---the one we call $U(1)_f$--- algebra is also
supposed to have only integer charge
representations, so that it is indeed
a representation of the {\em group} U(1) and not only of the
algebra or of the covering group $ \bf R$.
Actually it turns out, from our fit to
the quark-lepton mass spectrum described in section 5,
that the Higgs fields we introduce into our model
have somewhat strangely quantised $Q_f$ charges---the
abelian $U(1)_f$ charges. The  Higgs fields $Q_f$ charges
are quantised in units three times
smaller than those for the
quarks and leptons.

The original motivation for our non-simple anti-grand unified
group was inspired by the
ideas \cite{book,db2,brene,vancouver} of ``confusion'',
but we now think a more
convincing argument
is the one which we present below in section 4. Also we have
noted that the Standard Model {\em group}
$SMG = S(U(2)\times U(3))$ has a remarkably
low number of automorphisms for its
size. Then one might speculate that this
result is connected with groups
tending to get fewer and fewer automorphisms the more they
break down by the Higgsing. Such
argumentation also points, to some extent,
in the direction of our proposed group \cite{book,brene,brene2}.

\hspace{-0.5 cm}{\bf Desert almost to the Planck scale.}

It should
be stressed that it is part of our model that there
be essentially no new physics, except
for the Weinberg-Salam Higgs particle,
until an order of magnitude or so below the Planck energy scale.
That is to say that there shall for instance be no supersymmetric
partners before this scale, also no leptoquarks or the like.
When we say essentially it means
that some new particles could perhaps
be tolerated in our model provided
they do not disturb our predictions.
We namely get good predictions without
anything new at any accessible scale.

\hspace{-0.5 cm}{\bf Physically existing regularization.}

Although
quantum field theories are exceedingly successful, it has been
shown, for example, that scalar
self-interacting theory in 4 dimensions is
only consistent over an infinity of
scales provided the self-coupling
vanishes (the triviality bound). Presumably
most quantum field theories are
indeed inconsistent, unless they are
trivial in the sense of having
zero couplings, or asymptotically free.
So we can really not easily
have field theory without having some new physics
beyond the Standard Model that would appear much like a
regularization. Especially with gravity
becoming strong quantum mechanically
at the Planck scale, it would seem
strange if indeed there should be no new
physics---presumably being finite,  so that it could
appear as a regularization---at the Planck scale (or below).

In section 6 we shall describe computations
made as if the regularization was
provided by a real physically existing space-time
lattice, but we hope it does not matter too much what
form of regularization we use \cite{Larisa}.

\hspace{-0.5 cm}{\bf Order unity fundamental Yukawa couplings.}

The assumption
that the various fundamental Yukawa couplings are all of order unity
could perhaps have some rationale in our MPP principle, but
that would seem a rather stupid way to argue for it. Really it
is most natural to simply assume that any couplings, not having
a reason to be suppressed or to take on
special values, are unity order of magnitudewise.

\vspace{0.5 cm}

To these assumptions, we add some details about which
Higgs fields might
be responsible for the needed breakdown
of the gauge group. Also we make some assumptions, which in
principle can be checked by lattice calculations, as to
the phases which are supposed to meet at the multiple
point.
In particular the rather complicated choice of quantum numbers
for the Higgs fields $W$, $T$, $\xi$ and $S$, responsible for the
breaking of our group to
the Standard Model group, is really made as
a kind of discrete number fitting.
In fact it took some time to find a proposal yielding
a successful phenomenology.

We would like to stress that it is not
essential for all the features of
our model to be correct.
Even if part of
it turns out wrong, it does not necessarily fall apart completely.
Really the point is that to produce our results for the Higgs
particle and top quark masses
(see section 2), we only use the MPP assumption among our two
``basic'' assumptions.
For the ratios of the quark and
lepton masses and the quark mixing angles
(see section 5), for which we get
order of magnitude fits, we only
use the other basic assumption, the
anti-GUT gauge group.  For the fine structure constant predictions
(see section 6), however,
both basic assumptions are needed.
For the fine-tuning questions as to
why the cosmological constant and the theta angles
for SU(3) and SU(2) are zero, we need, if
we can do anything at all, only
the MPP assumption, but not the gauge group
beyond the Standard Model (see section 3).
The most difficult and mysterious
parameter---the scale of the weak interaction compared
to say the Planck scale---is not
yet predicted by our model, and it remains
a mystery why this scale
is so exceedingly low.

\section{Higgs Mass $135\pm 9$ GeV and Top Quark Mass $173\pm 5$ GeV}

\begin{figure}
\leavevmode
\vspace{-0.5cm}
\centerline{
\epsfig{file=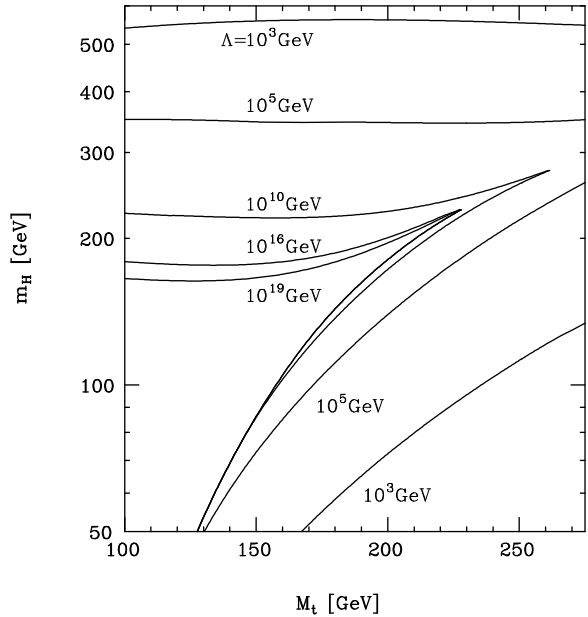,width=10.0cm,angle=90}}
\vspace{-1.0cm}
\caption{SM bounds in the ($M_t$,$m_H$) plane
for  various values of $\Lambda$, the scale at which
new physics enters. The lower part of each curve
coincides with the vacuum stability bound.
The upper part of each curve
corresponds to the so-called triviality bound
and the  point in the top
right-hand corner, where it meets the vacuum stability curve,
is the infra-red quasi-fixed point.}
\label{fig:Maiani}
\end{figure}

The application of the MPP to the pure Standard Model (SM) implies
that the SM parameters should be adjusted, such that there exists
another vacuum state degenerate in energy density with the
vacuum in which we live. This means that the effective SM
Higgs potential $V_{eff}(|\phi|)$
should, for example, have a second minimum
degenerate with the well-known
minimum at the electroweak scale $\langle |\phi| \rangle = 246$ GeV.
Thus we predict that our vacuum is barely stable and we
just lie on the
vacuum stability curve; the Higgs particle mass is predicted
to take on its lowest allowed value before our vacuum becomes
unstable. The form of the SM vacuum
stability curve in the top quark mass,
Higgs particle mass ($M_t, m_H$) plane depends on the physical
cut-off scale $\Lambda$ beyond which
the Standard Model is replaced by a
more fundamental theory, as illustrated \cite{zwirner}
in figure \ref{fig:Maiani}.
Taking the cut-off scale to be given by the Planck mass
$\Lambda \simeq M_{Planck} \simeq 10^{19}$ GeV
and the top quark mass from
experiment \cite{tipton}, $M_t = 175 \pm 6$ GeV, we
obtain the prediction:
\begin{equation}
M_H = 139 \pm 16 \; \mbox{GeV}
\label{eqmhiggs}
\end{equation}
for the Higgs particle pole mass.

However if the vacuum degeneracy requirement
should have a good chance of being physically relevant, the
range of $|\phi|$ values, lying between the vacuum expectation
values (VEVs) $|\phi_{vac\; 1}|$ and $|\phi_{vac\; 2}|$ for the two
vacua, should be as large as possible. This condition is
analogous to the necessity for an appropriately
large latent heat of fusion for water so that
slush---partially melted snow or ice---is likely
to be found in winter. That is to say we
have the strong first order phase transition requirement:
\begin{equation}
|\phi_{vac\; 2}| - |\phi_{vac\; 1}| \simeq \Lambda  \simeq M_{Planck}
\label{eqfirstorder}
\end{equation}
Since $|\phi_{vac\; 1}| = 246$ GeV, this requires
the second vacuum to have
a VEV of the order $|\phi_{vac\; 2}| \simeq M_{Planck}$.
If we now impose both the degenerate vacua and first order
phase transition requirements, we determine a single point
on the vacuum stability curve. In this way
we obtain \cite{smtop}, within
the pure Standard Model, predictions
for both the top quark and Higgs boson
pole masses:
\begin{equation}
M_{t} = 173 \pm 5\ \mbox{GeV} \quad M_{H} = 135 \pm 9\ \mbox{GeV}
\end{equation}
There is remarkably good agreement with the experimental
value of the top quark mass.

The above results were obtained by studying the renormalisation
group improved SM Higgs potential. Including quantum
fluctuations, the classical potential picks up loop corrections
and we get the effective Higgs potential \cite{sher}:
\begin{eqnarray}
\lefteqn{V_{eff}(|\phi|) = \frac{1}{2}m_{0H}^2 |\phi |^2
+  \frac{1}{8}\lambda_0 |\phi |^4
+ \frac{1}{2}\mbox{Tr}\,\log\,[\Delta^{-1}_{\mbox{Bosons}}(\phi)]
}\nonumber \\
 & & -{} \frac{1}{2}\mbox{Tr}\,\log\,[\Delta^{-1}_{\mbox{Fermions}}(\phi)]
+ \mbox{Higher order loop terms}
\label{veff}
\end{eqnarray}
Here $\Delta^{-1}_{\mbox{Bosons}}(\phi)$ and
$\Delta^{-1}_{\mbox{Fermions}}(\phi)$
denote the (appropriately normalised) inverse propagators
for boson and fermion fields in the background
Higgs field $\phi$. The parameters $m_{0H}^2$ and
$\lambda_0$ are the bare Higgs mass and bare Higgs
self-coupling constant respectively. An efficient way of
taking the higher order loops into account is to make use
of the running coupling constants---by far the most
relevant for our work is the Higgs field
self-interaction coupling
$\lambda(\mu)$---as calculated  by integrating the renormalisation
group equations for the various couplings
in the Standard Model. Then, by
identifying the scale parameter $\mu$ with the field value
$|\phi|$, the perturbative expansion is reorganised so
that the leading-log contributions from the loop
corrections are transferred to the tree level part of
the effective potential. The next to leading-log
terms can be included by using the
one-loop effective potential, with running couplings
evaluated using the two-loop renormalisation
group beta functions.

For the purposes of our discussion
it is sufficient to consider the renormalisation group improved
tree level effective potential:
\begin{equation}
V_{eff}(\phi)  =
\frac{1}{2}m_{H}^2(\mu = |\phi |) |\phi |^2
+ \frac{1}{8}\lambda (\mu = |\phi | )\, |\phi |^4
\end{equation}
In order that $|\phi_{vac\; 1}| = 246$ GeV, the
coefficient $m_{H}^2(\mu)$ has to be of the order
of the electroweak scale.
We are interested in values of the Higgs field
of the order $|\phi_{vac\; 2}| \simeq M_{Planck}$,
which is very large compared to the electroweak scale,
and for which the quartic term
strongly dominates the $\phi^2$ term;
so to a very good approximation
we have:
\begin{equation}
V_{eff}(\phi) \simeq
\frac{1}{8}\lambda (\mu = |\phi | ) |\phi |^4
\end{equation}
The running Higgs self-coupling constant $\lambda (\mu)$
is readily computed by means of the
renormalisation group equation:
\begin{equation}
\frac{d\lambda}{d\ln\mu} =
\beta_{\lambda}(\lambda,g_t,g_1,g_2,g_3)
\label{rgelam}
\end{equation}
Here the $g_i(\mu)$ are the three SM
running gauge coupling constants,
discussed further in section~\ref{sec:fine},
and $g_t(\mu)$ is the top quark running Yukawa coupling constant,
which satisfies the renormalisation group equation:
\begin{equation}
\frac{dg_t}{d\ln\mu} =
\beta_{g_t}(\lambda, g_t,g_1,g_2,g_3)
\label{rgetop}
\end{equation}
The beta functions are given to first order by:
\begin{equation}
16\pi^2\beta_{\lambda} =12\lambda^2 +
3\left(4g_t^2 - 3g_2^2 - g_1^2\right)\lambda
 + \frac{9}{4}g_2^4 + \frac{3}{2}g_2^2g_1^2
+ \frac{3}{4}g_1^4 - 12g_t^4
\label{betalam}
\end{equation}
and
\begin{equation}
16\pi^2\beta_{g_t} = g_t\left(\frac{9}{2}g_t^2 - 8g_3^2
- \frac{9}{4}g_2^2 - \frac{17}{12}g_1^2\right)
\label{betatop}
\end{equation}
The renormalisation group equations are
in practice solved numerically, using the second order
expressions \cite{drtjones} for the beta functions.

\begin{figure}
\leavevmode
\centerline{
\epsfig{file=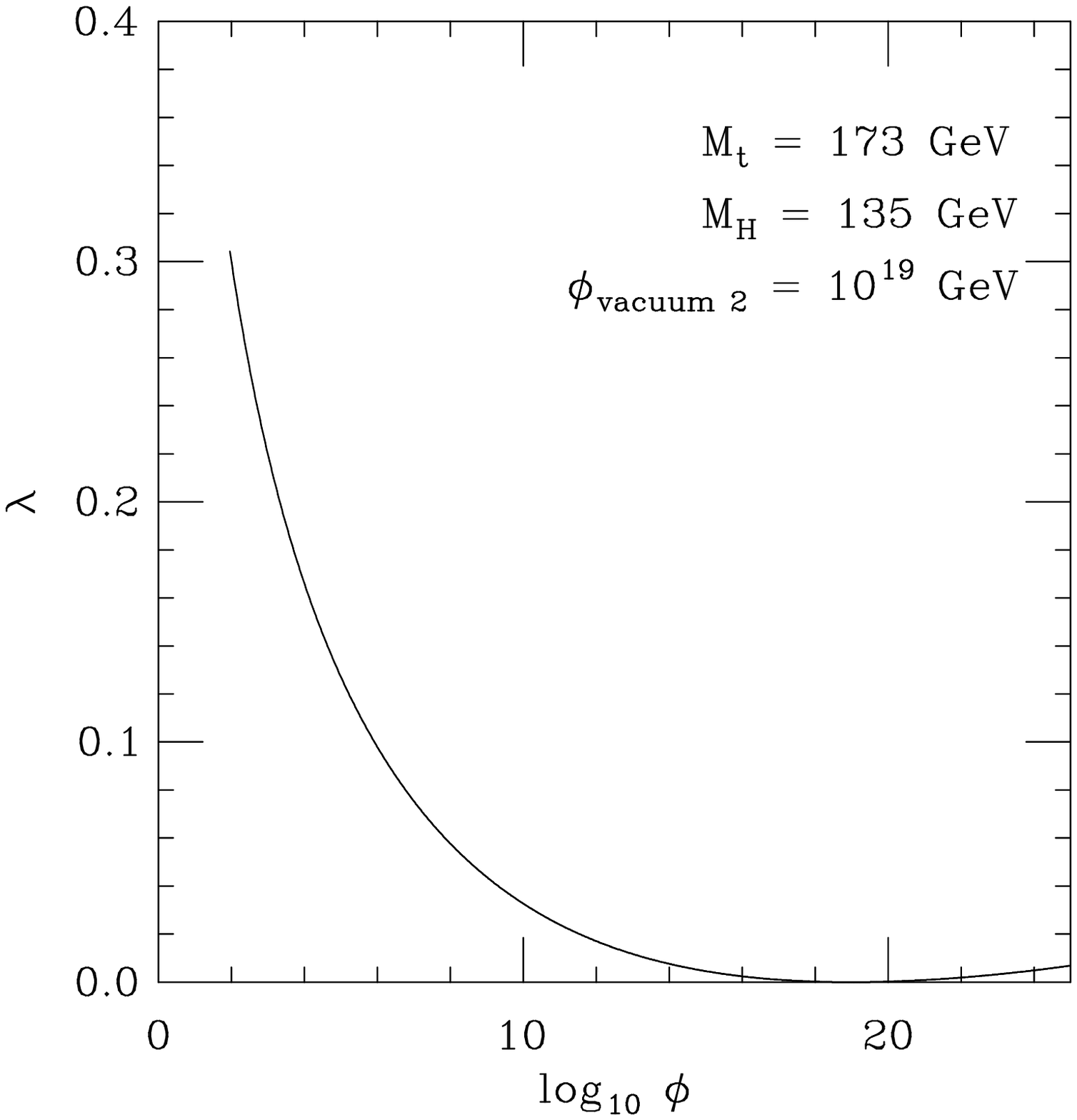,width=6.8cm}
\hspace{-0.6cm}
\epsfig{file=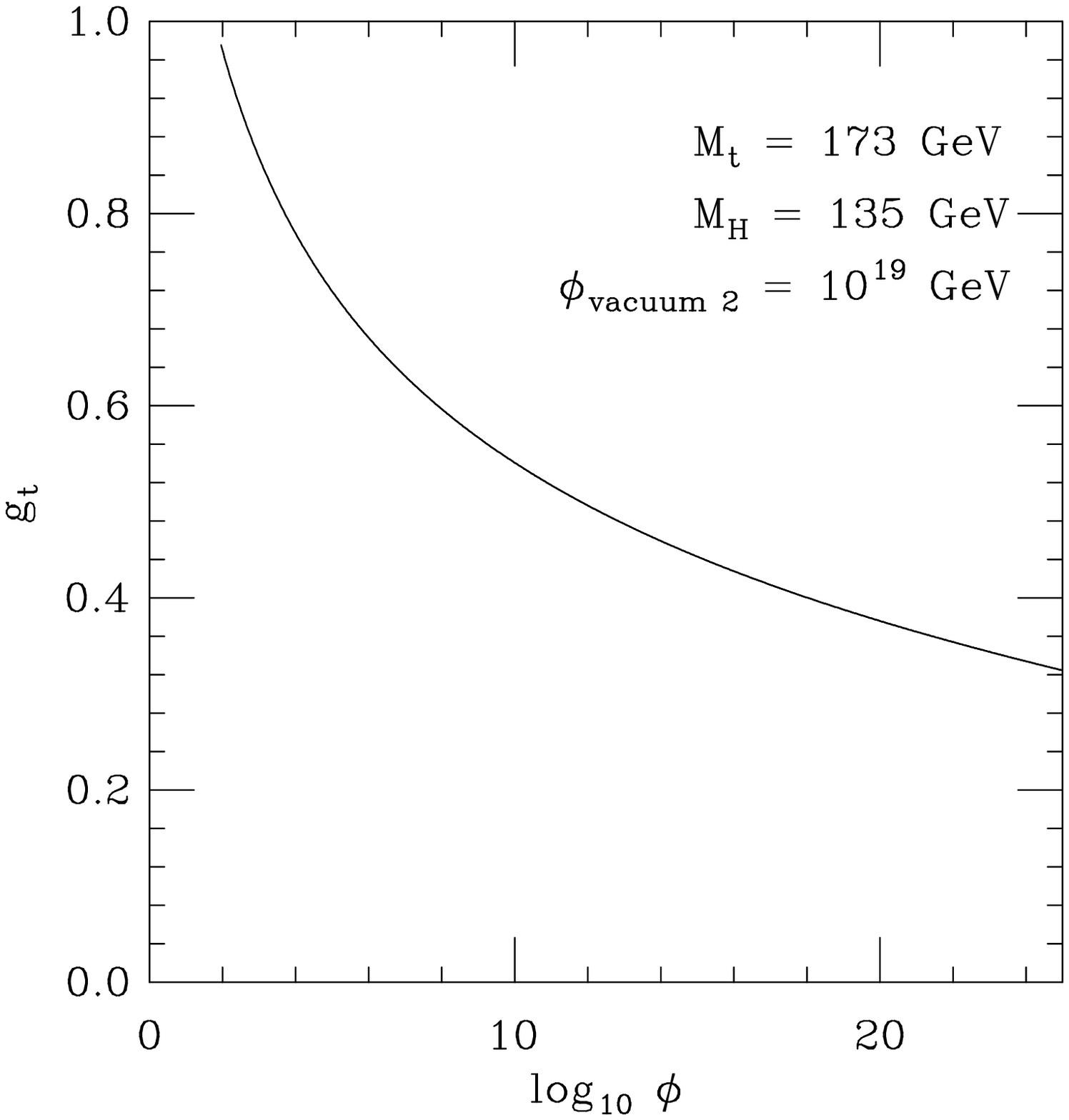,width=6.8cm}
}
\vspace{-0.6cm}
\caption{Plots of $\lambda$
and $g_t$ as functions
of the scale of the
Higgs field $\phi$ for degenerate vacua with the second Higgs
VEV at the Planck scale $\phi_{vac\;2}=10^{19}$ GeV.
We formally apply the second order SM renormalisation
group equations up to a scale of $10^{25}$ GeV.}
\label{fig:lam19}
\end{figure}

The vacuum degeneracy condition is imposed by requiring:
\begin{equation}
V_{eff}(\phi_{vac\; 1}) = V_{eff}(\phi_{vac\; 2})
\label{eqdeg}
\end{equation}
Now the energy density in vacuum 1 is exceedingly small
compared to $\phi_{vac\; 2}^4 \simeq M_{Planck}^4$. So
we basically get the degeneracy condition, eq.~(\ref{eqdeg}),
to mean that the coefficient $\lambda(\phi_{vac\; 2})$
of $\phi_{vac\; 2}^4$ must be zero with high accuracy:
\begin{equation}
\lambda(\phi_{vac\; 2}) = 0
\end{equation}
At the same $\phi$-value the derivative of the effective
potential $V_{eff}(\phi)$ should be zero, because it has
a minimum there. In the approximation $ V_{eff}(\phi) \approx
\frac{1}{8}\lambda(\phi) \phi^4 $ the derivative of $V_{eff}(\phi)$
with respect to $\phi$ becomes
\begin{equation}
\frac{dV_{eff}}{d\phi}|_{\phi_{vac \; 2}}
= \frac{1}{2}\lambda(\phi)\phi^3
+\frac{1}{8}\frac{d\lambda(\phi)}{d\phi}\phi^4
 = \frac{1}{8}\beta_{\lambda} \phi^3
\end{equation}
Thus at the second minimum of the effective potential we have:
\begin{equation}
\beta_{\lambda}(\mu = \phi_{vac\; 2}) = \lambda(\phi_{vac\; 2}) = 0
\end{equation}
which gives to leading order, setting $\lambda = 0$ in
eq. (\ref{betalam}), the relationship:
\begin{equation}
\frac{9}{4}g_2^4 + \frac{3}{2}g_2^2g_1^2 + \frac{3}{4}g_1^4 - 12g_t^4 = 0
\end{equation}
between the top quark Yukawa coupling and the gauge coupling
constants at the scale $\mu = \phi_{vac\; 2} \simeq M_{Planck}$.
We use the renormalisation group equations to relate the couplings
at the Planck scale to their values
at the electroweak scale.
Figure \ref{fig:lam19} shows
the running coupling constants $\lambda(\phi)$ and $g_t(\phi)$
as functions of $\log(\phi)$. Their values at the
electroweak scale give our predicted combination of pole
masses: $M_t = 173$ GeV and $M_H = 135$ GeV.

\begin{figure}
\leavevmode
\vspace{-0.5cm}
\centerline{\epsfig{file=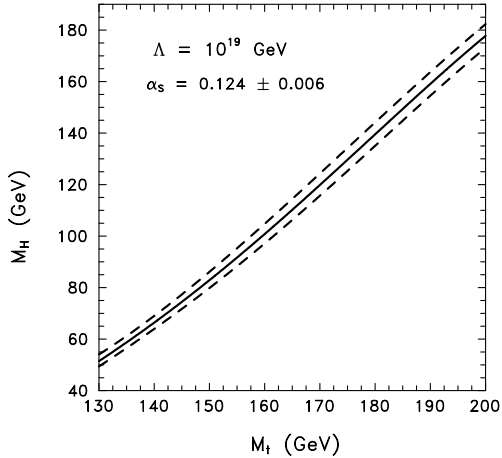,width=7cm,%
bbllx=95pt,bblly=130pt,bburx=510pt,bbury=555pt,%
angle=270,clip=}
}
\caption{SM vacuum stability curve
for $\Lambda = 10^{19}$ GeV and $\alpha_s = 0.124$ (solid line),
$\alpha_s = 0.118$ (upper dashed line), $\alpha_s = 0.130$
(lower dashed line).}
\label{fig:vacstab}
\end{figure}

The vacuum stability curve has been studied for the
Standard Model by several authors
\cite{shervs,isidori,casas}.
Their results are slightly different but, within errors,
are each consistent with the linear fit
\begin{equation}
M_H = 135 + 2 (M_t -173) - 4 \frac{\alpha_3 - 0.117}{0.006}
\label{eqvacstab}
\end{equation}
to the vacuum stability curve, in GeV units.
This is illustrated by the results of Casas {\em et al.} \cite{casas} in
figure \ref{fig:vacstab}.
When this degenerate minima condition eq.~(\ref{eqvacstab})
is combined with the experimental value
of the top quark pole mass, $M_t = 175 \pm 6$ GeV,
we obtain a rather clean MPP prediction for the Higgs
pole mass: $M_H = 139 \pm 16$ GeV.

If we now also impose the strong first order phase transition
requirement, which takes the form
$|\phi_{vac\; 2}| \simeq M_{Planck}$,
we no longer need the experimental top quark mass as an input,
but rather obtain our prediction for both $M_H$ and $M_t$.
A change in the scale
of the minimum $\phi_{vac \; 2}$ by an order of magnitude, from
$10^{19}$ GeV to  $ 10^{18}$ or $ 10^{20}$ GeV, gives a shift
in the top quark mass of about
2.5 GeV. Since the concept of Planck units only makes
physical sense w.r.t.\ order of magnitudes,
this means that we cannot,
without new assumptions, get a
more accurate prediction than of this
order of magnitude of 2.5 GeV
uncertainty in $ M_t$ and 5 GeV in $M_H$.
The uncertainty in the strong fine structure constant
$\alpha_3(M_Z)= 0.117 \pm 0.006 $ leads to an uncertainty in
our predictions of approximately $\pm$ 2\%,
meaning $\pm$
3.5 GeV in the top quark mass. So our overall
result for the top quark mass
is $ M_t = 173 \pm 5$ GeV.
Combining the
uncertainty from the Planck scale only being known in order of
magnitude and the $\alpha_3
$ uncertainty with the calculational uncertainty
in the vacuum stability curve, we get an overall uncertainty in
the Higgs boson mass of $\pm$ 9 GeV. So our Standard Model
multiple point prediction for both the top quark and
Higgs boson pole masses is:
\begin{equation}
M_{t} = 173 \pm 5\ \mbox{GeV} \quad M_{H} = 135 \pm 9\ \mbox{GeV}
\end{equation}

\section{Some Well-Known Fine-Tuning Problems}

By these well-known fine-tuning problems, we allude to the
questions as to
why the cosmological constant is zero
and why $\Theta_{QCD}$ and presumably also
the weak SU(2) topological
term coefficient $\Theta_{weak}$ are zero.
The gauge hierarchy problem
of why the W's and $Z^0$ are so light compared to the unified or
Planck scale also belongs to this class of fine-tuning problems.
However we do not
{\em yet} even have a partial solution
to the gauge hierarchy problem.

We can formally bring the cosmological
constant being zero and our
MPP principle together under the single postulate:
All vacua have energy density zero.

As far as we can think of our MPP as saying that the
Euclideanised theory
has a phase transition just at the parameters chosen by Nature,
we can make use of some lattice gravity calculations
by Ambj{\o}rn and Varsted and by Hamber \cite{ambjorn}.
These show that there is a phase transition
in the space of lattice gravity parameters,
where the cosmological constant is
zero. It is not so surprising that there should be a phase border
where the cosmological constant is zero, since
for instance the four-volume---of a universe development---grows to
infinity just when the effective
cosmological constant passes through zero.

Another similar fine-tuning problem to
the cosmological constant one
is the problem of why the theta-vacuum
parameter or topological term coefficient
$\Theta$ vanishes. Is it also thinkable
that this $\Theta$ is just zero
on some phase border?
In a calculation by Schierholz \cite{schierholz}
one does find a phase transition at the
$\Theta_{QCD} = 0$ hypersurface.

\section{The ``Anti-GUT'' Gauge Group $SMG^3\times U(1)_f$}

As mentioned in the introduction, the second major assumption
in our model is
that the gauge group $G$ to be found beyond the Standard Model,
close to the Planck scale,
should be $SMG^3\times U(1)_f$.
This group breaks down by the
Higgs mechanism to the SMG as the diagonal subgroup of its $SMG^3$
subgroup. This group
$SMG^3\times U(1)_f$, with its 3 x 12 + 1 = 37
generators, would at first seem a very arbitrary choice of
``unified group''. However, we shall actually now argue
it can be
characterized by postulating 4 to 5 not so unreasonable,
nor arbitrary assumptions about the gauge group $G$ beyond the
Standard Model (SM).

As a zeroth postulate, of course, our characterization should
have the property that the gauge group $G$
beyond the Standard Model must contain the Standard Model
Gauge group $SMG$ as a subgroup. In addition it should obey the
following 4 postulates---for which we also deliver some
hand-waving arguments:

$\bullet$  $G\subseteq U(45) $, the group of unitary
transformations of the known 45 Weyl fermions in the Standard Model.

This assumption is really the postulate that we
{\em totally} ignore those generators in the gauge group which
do not couple to the already known fermions in the Standard Model.
At first, it therefore just means that we
only look for a factor group $G$ of the subgroup
of the full group not transforming the known fermions
into yet to be found ones.
In fact $G$ should be that factor group which is
obtained by dividing out the subgroup leaving the
known fermions untouched.
However, even a part of the gauge group not coupling to the
known fermions could influence the interaction of the various
Higgs fields in our model below and thus influence our
phenomenology---and perhaps more importantly it could influence our
gauge coupling calculations in section 6.
Thus our restriction to $G\subseteq U(45)$
involves a physical  assumption about the theory and
is not simply a convenient definition of $G$ as the
factor group, which couples to the SM fermions,
of a presumably bigger group.
We may hope though that it is a good approximation for
phenomenological applications to ignore the rest of the true group.

In the spirit of using our assumption about many degenerate vacua,
one could cook up an argument for why
we should already know the fermions
to which the gauge fields couple:
If there are many phases with the same
zero temperature vacuum energy density, then those with
the highest number of ``light'' particle species (i.~e.
particles that can form
a Planck radiation immediately after the Big Bang)
will tend to push the other phases aside and come to dominate
the early Universe.
We would therefore expect the phase, surviving after the
Big Bang, to have
a maximal number of particle species with mass lighter than the scale
of mass relevant for the survival of phases. This would favour the
dominance, after the Big Bang, of a phase
with exceptionally many of the fermion
species remaining mass protected (chiral) after the breakdown of the
gauge group $G$ to the $SMG$---the gauge group in the present day
phase of the Universe.
But, in such a case,
we may still see a very large proportion of the fermion species
which existed at the outset of the Universe and, thus,
our assumption that all the gauge
fields belonging to $G$ couple to some known fermions
has an increased reliability.

$\bullet$ No anomalies.

There should neither be gauge nor mixed anomalies (nor discrete
anomalies, but it is not relevant here). This is an almost unavoidable
assumption, in as far as the
existence of anomalies would spoil the gauge symmetries.
Rather our real assumption is that we do {\em not} allow for a
Green-Schwarz \cite{green-schwarz} type of anomaly cancellation
to play any role in our model.
In the Standard Model itself there
is, of course, no need for any Green-Schwarz
anomaly cancellation.
This may be taken as
some support from phenomenology
for our assumption that only
rather straightforward anomaly cancellations take place.

$\bullet$ Keep irreducible representations of the Standard Model Group
irreducible under the big group $G$.

Grand Unified Theories, like $SU(5)$,
combine the SM irreducible representations
into larger GUT irreducible representations and thereby obtain
symmetry relations between the SM Yukawa couplings.
However the exact $SU(5)$ GUT degeneracies of dsb-quark and
charged lepton masses at the unification scale are really
not wanted, except for the case of $\tau$ being degenerate with $b$.
In fact the unwanted SU(5) predictions,
$m_{\mu} = m_s$ and $m_e = m_d$,
can only be tolerated
by having e.g. Georgi-Jarlskog factors \cite{georgijarlskog}
of 3 coming in, by postulating
several different Higgs representations to provide the various
quark and lepton masses. It would really be an advantage for the
GUT model agreement with data if one, as in our model, could
replace these
exact SU(5) predictions by {\em only order of magnitude
degeneracy predictions}.

Taking a crude and unbiased look at
the spectrum of quark and leptons, its most remarkable
feature is that almost every mass has a value deviating
by big factors from almost all the other masses.
At first sight, there does not seem to be even
degeneracy order of magnitudewise.
If that were the case---which is not really
true even in our model---it would suggest
that every irreducible representation should have
its own set of approximately conserved quantum number
combinations, so that almost no
degeneracies even order of magnitudewise should be likely to
occur. Thus the best way of ensuring that no degeneracy is
realised is to keep each irreducible representation
under the $SMG$ in a separate irreducible representation of $G$.

$\bullet$ Maximal G with the above constraints.

In order to obtain the order of magnitude mass splittings
mentioned above, we need as many as possible partially
conserved quantum numbers to provide different
suppressions for each mass term. If we assume, as we do,
that these quantum numbers are gauge quantum numbers, we
need as big a gauge group as possible. One might
also try to justify the maximal $G$ assumption
by postulating that, at
the Planck level, all the fields that
can couple to the Weyl fermions
and that are allowed will exist
as dynamical degrees of freedom \cite{rugh}.
For each irreducible n-dimensional representation
of the Standard Model, we could postulate
a priori a U(n) gauge group just transforming its n components.

\vspace{0.5 cm}

With these four postulates a somewhat cumbersome
calculation shows that,
modulo permutations of the various irreducible representations
in the Standard Model
fermion system, we are led to our gauge group
$SMG^3\times U(1)_f$.
Furthermore it shows that the $SMG$ is embedded
as the diagonal subgroup, as in our AGUT model.
We first consider the non-abelian part of the
sought-after group $G$. There are 5 x 3 = 15
irreducible representations of Weyl
fermions in the Standard Model,
classifying the 45 Weyl components into 3
sets with six components in
each (the left-handed quarks), 6 sets
with 3 in each (the right-handed
quarks), 3  sets with 2 in each (the left-handed leptons), and
finally the 3 lonely components (the right-handed leptons).
Our first postulate means that we only consider
transformations of these 45 Weyl components.
The third postulate means that we do not allow those
transformations in $G$, which transform
components in one of the mentioned
15 sets of components into another one.
So $G$ must be a subgroup of the group of all the
transformations inside these 15 sets which do not mix
them: $U(6)^3 \times U(3)^6 \times U(2)^3 \times U(1)^3$.
Imposing our second postulate, we must avoid anomalies
and thus make some identifications among the $SU(m)$'s
and $U(1)$'s contained in the various $U(n)$ groups.
However in order to make the resulting total group $G$
as big as possible, as our final postulate claims,
we should make as few identifications as possible.

It follows from our zeroth postulate ($SMG \subset G$) that the
transformations on the left-handed quarks cannot be
restricted more than to lie in the Standard Model. Consequently
it turns out that we cannot escape anomalies for the $SU(6)$'s.
So we must give up having any SU(6) and
must rather be satisfied with their $SU(2)\times SU(3)$ subgroups.
We must then identify
the $SU(2)$'s and $SU(3)$'s  in
the left-handed quark transformations with
those on the other sets of
components. The minimal amount of
identification of these non-abelian groups
turns out to correspond, up to permutations, to having
just three generations of quarks and leptons, each
with its own $SU(2)$ and $SU(3)$. At this stage of the
argument, apart from the right-handed leptons which
have no non-abelian groups acting on them, we have a
generation structure in the sense that
we have different SU(3)'s and SU(2)'s for
the three proto-generations.
However these proto-generations do not have to correspond
exactly to the experimentally observed
generations---in fact we shall see below that, in our fit,
we let the right-handed 2/3 charge
quarks, $c_R$ and $t_R$, be permuted
relative to the proto-generation structure.

The most complicated part of
the calculation is to decide which
identifications of the abelian groups
have to be made in order to avoid anomalies, i.e. how
big a subgroup of the $U(1)^{15}$  can avoid having anomalies
and be allowed in $G$.
In searching for the generators of an allowed subgroup, one
may expand them in terms of the generators for these
15 $U(1)$ 's and they have to obey some first order
(linear) relations for the coefficients,
in order to avoid anomalies
involving also the non-abelian or gravitational fields.
Also there are third order relations that have to be satisfied, in
order that there be no anomalies involving only
the subspace of abelian generators. It turns out that, with
the three generations of fermions, there are too many constraints to
be solved with an abelian subgroup of dimension higher than 4.
It is found that three of the allowed abelian generators in $G$
can be taken to be the 3 weak hypercharges, each defined to act
on only one generation. After that choice
the scheme becomes so tight that, apart from
various rewritings and permutations
of the particle names, there is a unique fourth $U(1)$ allowed
and that is what we call $U(1)_f$. Several of the anomalies
involving this $U(1)_f$ are cancelled by assigning
equal and opposite values of the $U(1)_f$ charge to
the analogous particles belonging to second and
third generations, while the
first generation particles have just zero charge \cite{davidson}.
The $U(1)_f$ can in fact be chosen to obey the following rules:

All members of the first generation carry zero $U(1)_f$ charge.

Left-handed particles, i.e.~doublets under the $SU(2)$ of
the Weinberg-Salam model, carry no $U(1)_f$ charge either.

The right-handed leptons and right-handed dsb-quarks in
the same proto-genera-tion carry the same $U(1)_f$ charge, as
if they obeyed an $SU(5)$ symmetry .

The $U(1)_f$ charge is opposite on the (right-handed)
2/3 electric charge quark and the -1/3 electric charge quark
in the same proto-generation.

Using these rules the $U(1)_f$ charges are totally given,
except for an overall normalisation and sign convention.
For example, we choose the right-handed b-quark,
and thus also the right-handed $\tau$ lepton, to have
the charge $Q_f= 1$. Then the proto-right-handed t-quark gets
the charge $Q_f = -1$. However we note that there is a
finesse of our fit to the fermion spectrum, according
to which the right-handed component of the experimentally
observed t-quark is actually the one having second
generation $SU(3)$ quantum numbers and is thus really
the proto-right-handed charm quark.
In a similar way the
right-handed component of the experimentally observed
charm quark has the third generation $SU(3)$
representation and is really the proto-right-handed top quark.
It is only the {\em right}-handed top and charm quarks that
are permuted in this way, while for example the left-handed
components are not.

\section{Masses and Mixing Angles for Quarks and Leptons}
\label{sec:masses}

We now consider the use of the gauge
group $ SMG^3\times U(1)_f$ breaking down
to the SMG, embedded as the diagonal subgroup
of the $SMG^3$, to make a fit \cite{smg3m}
to the orders of magnitude of the quark and lepton masses
and mixing angles (i.e.~the Cabbibo Kobayashi Maskawa matrix).
We propose a system of Higgs fields performing
the needed breaking to the SMG and
a candidate for being identified with
the Weinberg-Salam Higgs field of the Standard Model.
The idea is that the proposed Higgs fields have
expectation values which are small compared to
the fundamental scale (the Planck scale),
so that the gauge group charges broken by
these Higgs fields will be approximately
conserved. These partially conserved
quantum numbers can now take different values on
the right- and the left-handed components of a quark or
lepton, thereby making the mass term---or the effective SM Yukawa
coupling to the Weinberg-Salam Higgs field---forbidden
in first approximation. It is the main idea \cite{fn} that
this kind of forbiddenness is responsible for the
suppression of quark-lepton mass terms or, equivalently,
for their effective SM Yukawa couplings being smaller
than of order unity.

It is a significant assumption and the philosophy
of our model that all the fundamental Yukawa couplings---and
other couplings too---in our basic model are of order unity,
only the Higgs field expectation values can be small!
Since we have decided, for the mass matrices, only
to hope to predict---fit---the order of magnitudes
of the masses and mixing angles, it means that
we have simply set the Anti-GUT Yukawa couplings to
unity. However, in the computer calculation
of the mass spectrum we provide each mass matrix
element with a complex factor of order unity. We
use a random number generator to generate these factors,
with random phases and magnitudes within a factor of
two or three or so from unity. Finally we
average the output masses and mixing matrix
in a geometrical way (i.e. average the logarithms)
to obtain our predictions.
But really the order of magnitude results
can be estimated with reasonable accuracy by head,
because only one or at least very few terms
dominate a given quantity.
It is precisely our philosophy
that ``small'' Higgs expectation values
are very small, in the sense that
when you have a series of several products of them,
the largest term in the series
will be a good enough approximation to the sum of the series.

\begin{figure}
\begin{picture}(24000,11000)(-5600,0)
\THICKLINES

\drawline\fermion[\E\REG](0,1500)[6000]
\drawarrow[\E\ATBASE](\pmidx,\pmidy)
\global\advance \pmidy by -2000
\put(\pmidx,\pmidy){$b_L$}

\put(6000,0){$\lambda_1$}

\drawline\fermion[\E\REG](6000,1500)[6000]
\drawarrow[\E\ATBASE](\pmidx,\pmidy)
\global\advance \pmidy by -2000
\put(\pmidx,\pmidy){$M_F$}

\put(12000,0){$\lambda_2$}

\drawline\fermion[\E\REG](12000,1500)[6000]
\drawarrow[\E\ATBASE](\pmidx,\pmidy)
\global\advance \pmidy by -2000
\put(\pmidx,\pmidy){$M_F$}

\put(18000,0){$\lambda_3$}

\drawline\fermion[\E\REG](18000,1500)[6000]
\drawarrow[\E\ATBASE](\pmidx,\pmidy)
\global\advance \pmidy by -2000
\put(\pmidx,\pmidy){$b_R$}

\drawline\scalar[\N\REG](6000,1500)[4]
\global\advance \pmidx by 1000
\global\advance \pmidy by 1000
\put(\pmidx,\pmidy){$\Phi_{WS}$}
\global\advance \scalarbackx by -530
\global\advance \scalarbacky by -530
\drawline\fermion[\NE\REG](\scalarbackx,\scalarbacky)[1500]
\global\advance \scalarbacky by 1060
\drawline\fermion[\SE\REG](\scalarbackx,\scalarbacky)[1500]

\drawline\scalar[\N\REG](12000,1500)[4]
\global\advance \pmidx by 1000
\global\advance \pmidy by 1000
\put(\pmidx,\pmidy){$W$}
\global\advance \scalarbackx by -530
\global\advance \scalarbacky by -530
\drawline\fermion[\NE\REG](\scalarbackx,\scalarbacky)[1500]
\global\advance \scalarbacky by 1060
\drawline\fermion[\SE\REG](\scalarbackx,\scalarbacky)[1500]

\drawline\scalar[\N\REG](18000,1500)[4]
\global\advance \pmidx by 1000
\global\advance \pmidy by 1000
\put(\pmidx,\pmidy){$T$}
\global\advance \scalarbackx by -530
\global\advance \scalarbacky by -530
\drawline\fermion[\NE\REG](\scalarbackx,\scalarbacky)[1500]
\global\advance \scalarbacky by 1060
\drawline\fermion[\SE\REG](\scalarbackx,\scalarbacky)[1500]

\end{picture}
\caption{Feynman diagram for bottom quark mass in the full theory.
The crosses indicate the couplings of
the Higgs fields to the vacuum.}
\label{MbFull}
\end{figure}
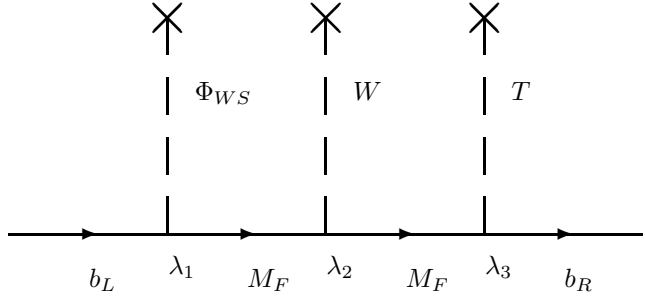

So we introduce
some Higgs fields, named $W$, $T$
and $\xi$, and the Weinberg-Salam Higgs field $\phi_{WS}$
with small vacuum expectation values
compared to the fundamental scale $M_{Planck}$.
As a result, whenever these Higgs fields are required to generate
mass terms, they cause suppressions of those terms.
In addition we introduce one Higgs field $S$ with a
vacuum expectation value (VEV)
of order unity on this scale.
We adjust the quantum numbers of these fields,
so as to make a fit of all the orders of magnitude of the
matrix elements in the three charged fermion mass matrices,
with products of the corresponding suppression factors.

The philosophy really is that there are a huge number of species of
vector-like Dirac fermions, with unsuppressed masses $M_F$
of the order of the fundamental mass $M_{Planck}$.
One can then form chain diagrams of the type shown
in figure \ref{MbFull},
in which the VEVs of the needed Higgs fields can, for example,
cause the effective b-quark mass matrix element to become non-zero.
For a given choice of the quantum numbers under
$SMG^3\times U(1)_f$ for the various Higgs fields,
we can estimate the orders of magnitude of
the various mass matrix elements. They are
given by products of the small numbers
denoting the VEVs in the fundamental units
of the fields $W$, $T$, $\xi$ and
the of order unity VEV of $S$. With the quantum number
choice that seems to fit the data,
we find the following orders of magnitude
for the effective SM Yukawa coupling matrix
elements---but remember that ``random'' order
unity factors are supposed to multiply all the matrix
elements---for the uct-quarks:
\begin{equation}
Y_U = \left ( \begin{array}{ccc}
	S^{\dagger}W^{\dagger}T^2(\xi^{\dagger})^2 & W^{\dagger}T^2\xi &
		(W^{\dagger})^2T\xi \\
	S^{\dagger}W^{\dagger}T^2(\xi^{\dagger})^3 & W^{\dagger}T^2 &
		(W^{\dagger})^2T \\
	S^{\dagger}(\xi^{\dagger})^3 & 1 & W^{\dagger}T^{\dagger}
			\end{array} \right ) \label{Y_U}
\end{equation}
the dsb-quarks:
\begin{equation}
Y_D = \left ( \begin{array}{ccc}
	SW(T^{\dagger})^2\xi^2 & W(T^{\dagger})^2\xi & T^3\xi \\
	SW(T^{\dagger})^2\xi & W(T^{\dagger})^2 & T^3 \\
	SW^2(T^{\dagger})^4\xi & W^2(T^{\dagger})^4 & WT
			\end{array} \right ) \label{Y_D}
\end{equation}
and the charged leptons:
\begin{equation}
Y_E = \left ( \hspace{-0.2 cm}\begin{array}{ccc}
	SW(T^{\dagger})^2\xi^2 & W(T^{\dagger})^2(\xi^{\dagger})^3 &
		(S^{\dagger})^2WT^4\xi^{\dagger} \\
	SW(T^{\dagger})^2\xi^5 & W(T^{\dagger})^2 &
	(S^{\dagger})^2WT^4\xi^2 \\
	S^3W(T^{\dagger})^5\xi^3 & (W^{\dagger})^2T^4 & WT
			\end{array} \hspace{-0.2 cm}\right ) \label{Y_E}
\end{equation}

These Yukawa
matrices were calculated using the
Higgs field abelian quantum numbers given below
in the form of U(1) charge vectors:
\begin{equation}
\vec{Q} \equiv \left ( \frac{y_1}{2},\frac{y_2}{2},
\frac{y_3}{2},Q_f \right )
\end{equation}
We let the non-abelian representations
be consequences of the specified abelian quantum numbers,
by imposing the natural generalisation of the
Standard Model charge quantisation rule:
\begin{equation}
y_i/2 + d_i /2 + t_i/3 = 0 \quad ( \mbox{mod} \quad 1)
\end{equation}
We also require that
the non-abelian representations be the smallest
possible with the dualities $d_i$
and/or trialities $t_i$ determined from the quantisation rule.
The abelian charge vectors are as follows:
\begin{equation}
\vec{Q}_{\phi_{WS}} = ( 0,2/3,-1/6,1)
\end{equation}
\begin{equation}
\vec{Q}_W  =  ( 0,-1/2,1/2,-4/3) \quad
\vec{Q}_T  =  ( 0,-1/6,1/6 , -2/3)
\end{equation}
\begin{equation}
\vec{Q}_{\xi}  =  ( 1/6,-1/6,0,0) \quad
\vec{Q}_S  =  ( 1/6,-1/6, 0 ,-1)
\end{equation}

These quantum numbers were chosen with
a view to fitting the
mass and mixing angle data extrapolated to the Planck
scale, using the SM renormalisation group equations for
$Y_U$, $Y_D$ and $Y_E$.
Actually we have chosen the Weinberg-Salam Higgs field
quantum numbers in our gauge group to be such that the
mass matrix element suggested to give the top
quark mass---namely the transition between the
formally generation number 2
right-handed charge 2/3 Weyl field ``$c_R$''
and the left-handed third generation
field $t_L$---is unsuppressed. In other words,
we simply took the quantum numbers for the
$\phi_{WS}$ field to be the difference between
those of these two Weyl fields.

Next we studied
the predictions that follow from our gauge group
quantum number pattern, but are not sensitive to the
choice of Higgs field quantum numbers.
We can for instance find some inequalities, but
the most important result is that
corresponding proto-diagonal
elements in each of the Yukawa matrices $Y_U$, $Y_D$ and $Y_E$
have the same  3 quantum number differences.
This feature can, for example with our choice of
Higgs field quantum numbers, be seen in the matrices of
eqs.~(\ref{Y_U}-\ref{Y_E}). Along the diagonals of
these three matrices, the suppression factors are given by
$S W (T^{\dagger})^2  \xi^2 $,
$ W (T^{\dagger})^2$ and $ W T$. Well, in the
uct-matrix it is the hermitean conjugated fields that occur
along the diagonal, but it does not matter,
because the suppression is the same anyway. It implies that,
{\em if} the proto-diagonal dominates the eigenvalues,
the masses within a generation will be order of
magnitudewise degenerate.
So, for example,
the $u$, the $d$ and the electron are predicted to be
order of magnitudewise degenerate, without having
to specify the choice of Higgs quantum numbers,
provided only that it happens that these
proto-diagonal elements dominate.

Proto-diagonal dominance and order of magnitude
generation mass degeneracy work well
for the leptons being degenerate with the dsb-quarks
and for the whole first generation, i.e.~both
up and down quarks
being degenerate with the electron. But clearly the
top quark and the charm quark are not order of magnitudewise
degenerate within their generations! It is
therefore necessary, in our model, to arrange for the
top and the charm quarks to get the dominant
contributions to their masses from matrix elements not on the
proto-diagonal. But that then means the
right- and left-handed components of these
two quarks do not, at the proto-level,
belong to the same generation.
For example the left-handed component of the charm quark
couples to the second generation colour group,
while its right-handed component couples to the
third generation $SU(3)$ group (as well as third
generation weak hypercharge). It was
for this reason that we had to organize the top quark
mass to be given by a proto-off-diagonal matrix element, and
then it could be unsuppressed relative to
the electro-weak scale.

This rule of the same proto-diagonal in all
the three mass matrices gives most of the predictive
power for quark and lepton masses in our model.
So let us give an idea, as best we can, the reason
why there is such a rule:
It is well-known that the same Weinberg-Salam Higgs field
can provide masses to all the three types of charged fermions
in the simple Standard Model.
Now, with our $SMG^3$ group
the quantum number differences, between the right-
and left-handed fermions, on the proto-diagonal become
just the quantum numbers of the simple Weinberg-Salam
Higgs field, translated to belong to the generation
in question.
For instance, the quantum number difference needed to be provided to
give a mass to the first generation proto-diagonal matrix element
in, say, the dsb-quark mass matrix is
$\vec{Q} = (1/2 ,0 ,0 ,0 )$.
Using the charge quantisation rule and small representation rule
to translate the abelian quantum numbers into non-abelian
ones, this quantum number difference corresponds to no
colour but doublet under the $SU(2)$ of the first generation.
One namely gets the same quantum numbers as those of
the simple Weinberg-Salam Higgs doublet and $y/2$ = 1/2 but just
for the first generation, as long as we ask for
the first element along the diagonal.
We find, of course, the same quantum numbers in uct-quark and
charged lepton mass matrices. That is why we have this rule
of the same suppressions for the three proto-diagonals.

So for the $SMG^3$ group, we understand
this proto-diagonal quantum number difference rule,
but what about the $U(1)_f$? We can simply check
that, perhaps miraculously, the $U(1)_f$
quantum number differences along the proto-diagonals
also turn out to be the same in all the three mass matrices.
So the rule becomes general in our model!
It is this rule that simulates the GUT SU(5) mass predictions,
namely the degeneracy of the dsb-quarks with the charged leptons
in the corresponding generations. Note, however, that we {\em
only get the prediction of these degeneracies at the Planck scale
as far as order of magnitude is concerned, and not exactly!}
This gives much better agreement with experiment than exact
SU(5) predictions, which are rather bad unless more
Weinberg-Salam Higgs fields are included
a la Georgi-Jarlskog's factor 3 mechanism \cite{georgijarlskog}.
Also note that we {\em in addition predict that the up-quark
is degenerate with the down-quark and the electron}!
This does not follow
just from GUT SU(5), although the up-quark
is equally, not to say better,
degenerate with the electron than the down quark!

As well as the proto-diagonal quantum number difference rule, we
have the result that the quantum number difference on the
element responsible for the top quark mass is balanced
by $\phi_{WS}$.
Using these facts we could study various
relations between quantum numbers
and use experimental mass data to suggest Higgs fields and their
expectation values. In this way we found the system described by the
quantum numbers above.

{}From the Fritzsch rule \cite{Fritschrule}
for the mixing matrix element $V_{12} = \sin \theta_{Cabbibo}$
between the first and second generations,
it is suggested that the two off-diagonal
matrix elements connecting the
d-quark and the s-quark be equally big.
We take this to indicate that
these two elements
in the dsb-quark mass matrix should have essentially the same
approximately conserved quantum number differences. We achieve this
in our model by introducing a special Higgs field
$S$, with quantum numbers equal to the
difference between the quantum number
differences for these 2 matrix elements in the dsb-quark matrix.
Then we postulate that this Higgs field has
a VEV of order unity in fundamental units,
so that it does not cause any suppression but
rather ensures that the two matrix elements get equally suppressed.

The existence of a non-suppressing
field $S$ means that we cannot
control phenomenologically when this $S$-field is used.
Thus all the quantum numbers of the other
Higgs fields, found
by fitting data, can only have their
quantum numbers predicted modulo those
of the field $S$. We should therefore somehow, by requiring
small representations or the like, seek
to guess the best quantum numbers
for the fields $W$, $T$, $\xi$ and $\phi_{WS}$ by
adding adjustable multiples of
$\vec{Q}_S$.
The result of such a guess, using small
representations as the principle,
is provided by the following set of quantum numbers:
\begin{equation}
\vec{Q}_{\phi_{WS}} = (1/6 ,1/2 ,-1/6,0)
\end{equation}
\begin{equation}
\vec{Q}_W = (-1/6 ,-1/3,1/2,-1/3) \quad
\vec{Q}_T = (-1/6,0,1/6 , 1/3)
\end{equation}
\begin{equation}
\vec{Q}_{\xi} = (0,0,0,1) \quad
\vec{Q}_S = ( 1/6,-1/6, 0 ,-1)
\end{equation}
With this new pattern of quantum numbers
the powers of $S$ in the
mass matrices would be changed but, since
$S$ is 1 or at least close to 1,
it would not change the predictions significantly.

The VEVs of the Higgs fields $W$, $T$ and $\xi$
are numbers to be fitted in our model.
Thus we have only three
parameters to fit the nine quark and lepton
masses, the three mixing angles
and the CP-violation. In a way even the
overall scale of the masses is
predicted, since the top quark Yukawa coupling
is not suppressed and therefore of order unity.
So we get 9 + 4 numbers out of three
parameters, but only as far as order of magnitude
is concerned. To the extent that we only care for
orders of magnitude it should not matter that,
in the computer calculation, we make use
of complex random numbers of order unity to
average statistically the predictions from our order
of magnitude coupling matrices eqs. (\ref{Y_U}-\ref{Y_E}).
In principle, of course, there is
some parameter in the precise specification
of the random distribution.

\begin{table}
\begin{minipage}[t]{6.0cm}
\caption{Best fit to conventional experimental data.
All masses are running
masses at 1 GeV except the top quark mass which is the pole mass.}
\begin{displaymath}
\begin{array}{|ccc|}
\hline
 & {\rm Fitted} & {\rm Experimental} \\ \hline
m_u & 3.6 {\rm \; MeV} & 4 {\rm \; MeV} \\
m_d & 7.0 {\rm \; MeV} & 9 {\rm \; MeV} \\
m_e & 0.87 {\rm \; MeV} & 0.5 {\rm \; MeV} \\
m_c & 1.02 {\rm \; GeV} & 1.4 {\rm \; GeV} \\
m_s & 400 {\rm \; MeV} & 200 {\rm \; MeV} \\
m_{\mu} & 88 {\rm \; MeV} & 105 {\rm \; MeV} \\
M_t & 192 {\rm \; GeV} & 180 {\rm \; GeV} \\
m_b & 8.3 {\rm \; GeV} & 6.3 {\rm \; GeV} \\
m_{\tau} & 1.27 {\rm \; GeV} & 1.78 {\rm \; GeV} \\
V_{us} & 0.18 & 0.22 \\
V_{cb} & 0.018 & 0.041 \\
V_{ub} & 0.0039 & 0.0035 \\ \hline
\end{array}
\end{displaymath}
\label{convbestfit}
%
\end{minipage}
\hfill
\begin{minipage}[t]{6.0cm}
\caption{Best fit using alternative light quark masses extracted from
lattice QCD. All masses are running
masses at 1 GeV except the top quark mass which is the pole mass.}
\begin{displaymath}
\begin{array}{|ccc|}
\hline
 & {\rm Fitted} & {\rm Experimental} \\ \hline
m_u & 1.9 {\rm \; MeV} & 1.3 {\rm \; MeV} \\
m_d & 3.7 {\rm \; MeV} & 4.2 {\rm \; MeV} \\
m_e & 0.45 {\rm \; MeV} & 0.5 {\rm \; MeV} \\
m_c & 0.53 {\rm \; GeV} & 1.4 {\rm \; GeV} \\
m_s & 327 {\rm \; MeV} & 85 {\rm \; MeV} \\
m_{\mu} & 75 {\rm \; MeV} & 105 {\rm \; MeV} \\
M_t & 192 {\rm \; GeV} & 180 {\rm \; GeV} \\
m_b & 6.4 {\rm \; GeV} & 6.3 {\rm \; GeV} \\
m_{\tau} & 0.98 {\rm \; GeV} & 1.78 {\rm
\; GeV} \\
V_{us} & 0.15 & 0.22 \\
V_{cb} & 0.033 & 0.041 \\
V_{ub} & 0.0054 & 0.0035 \\ \hline
\end{array}
\end{displaymath}
\label{newbestfit}
\end{minipage}
\end{table}

The fit of the suppression factors or Higgs
VEVs is presented
in table \ref{convbestfit},
where we used the conventional quark masses \cite{PDG}.
In table \ref{newbestfit} we present a fit to the data
using recent lattice calculation estimates \cite{udsmasses}
of the current algebra masses
for the lighter quarks.
We performed this order of magnitude fit by minimising a
``$\chi^2$'' function, defined to be the sum of the
squares of the natural logarithms of the ratios of the
predicted to the ``experimental'' values of
the fermion masses and mixing matrix elements.
For order of magnitude fits, both fits agree very well with the
data. With the conventional quark masses the fitted
values of the suppression factors,
proportional to the Higgs field VEVs, become:
\begin{equation}
\langle W\rangle  =  0.179 \quad
\langle T\rangle  =  0.071  \quad
\langle \xi\rangle  =  0.099 \label{WTxivev}
\end{equation}
with $\chi^2=1.87$.
We here fitted the 9 masses and the 3 mixing angles
with the three parameters.
Thus our 12 - 3 = 9 predictions each on the average
deviate by a factor the squared logarithm of which
is 1.87/9 = 0.21, meaning $\sqrt{0.21} = 46 \%$
disagreement typically.
The values of the Higgs VEVs---or rather suppression
factors---for the
fit with the lattice quark masses are:
\begin{equation}
\langle W\rangle  =  0.123 \quad
\langle T\rangle  =  0.079  \quad
\langle \xi\rangle  =  0.077 \label{lattvev}
\end{equation}
and this fit has a larger value of $\chi^2 = 3.81$,
corresponding to $\sqrt{3.81/9}$ = $ 65 \% $ deviations.
But even this is good
for an order of magnitude fit.

In these fits we did not use the CP-violation
information as input, but we
predict from the fit the
CP-violating area of the ``unitarity triangles'' \cite{Jarlskog}
to be:
\begin{equation}
J \approx 5.8 \times 10 ^{-6} \qquad \mbox{for
conventional quark masses}
\end{equation}
and
\begin{equation}
J \approx 1.2 \times 10 ^{-5} \qquad \mbox{for lattice quark masses}
\end{equation}
The ``experimental'' value \cite{ExpCP}
derived from the observed CP-violation is
\begin{equation}
J \approx 2.0 \times 10^{-5} \quad \mbox{to} \quad  3.5 \times 10^{-5}
\end{equation}
Our prediction is about a factor 4 below the
data. However we should bear in mind that
the quantity $J$ is, in our model, a product of
very many factors and is
expected to be more uncertainly predicted than most
other quantities.

It is worthwhile to see the rather simple relations
one obtains from our model by
eliminating the suppression factors:
First one gets the already mentioned
degeneracy of the masses in the
same generation, except for the top and the charm
quarks (all after transport
by the renormalisation group to the Planck scale).
In addition we have the following order
of magnitude Planck scale relations:

a)   $ m_b^3$ = $ m_t m_c m_s $

b)    $V_{ub}$ = $ V_{td} $ = $ V_{12} V_{23}$

c )  $J$ (for CP-viol.)  = $V_{ub}  V_{12} V_{23}$

d)    $V_{23}= \frac{ m_s^2}{ m_c m_b }$

e)    $V_{12} = \sqrt{ \frac{ m_d}{m_s}}$

\section{Degenerate Vacua and Lattice Artifact Confinement;
\newline
the Fine-Structure Constants}
\label{sec:fine}

As we have assumed a regularization to truly exist and that
we could
take it to be lattice regularization without changing
our results concerning phase transitions too much,
we shall now think of phase
diagrams for lattice gauge theories. It is well-known that if
a lattice action for a lattice
gauge theory depends on a couple of
parameters, as in the example of a
non-abelian SU(2) Yang-Mills field with
both a trace of the doublet representation
and a trace of the triplet
representation of the plaquette variables
occurring in the action, then
the coefficients to these trace terms span a 2-dimensional
phase diagram in which three phases
arise \cite{Dashen}.
As we also stated above, the requirement
of degenerate vacua becomes, in the
Euclideanised formulation, the
requirement of various phases coming
together in the phase diagram space spanned by the
lattice action parameters.
With our very complicated non-simple Anti-GUT gauge group
$SMG^3\times U(1)_f$
a huge number of phases
are possible.

What we ideally should have done to evaluate the fine-structure
constants was the following:

We make a Monte Carlo computer simulation of
the lattice gauge theory for our favourite
gauge group $SMG^3\times U(1)_f$ using an
action with a rather large number of parameters.
There should, say, be terms with real
parts of traces of the many different
irreducible representations of the gauge group.
We then use Monte Carlo
methods to map out the phase diagram
by measuring various expectation values and observing
when they jump.
The phase diagram should reveal hypersurfaces separating
various phases that sometimes intersect
other such hypersurfaces along
new hypersurfaces of progressively higher co-dimension.
Now we look
for a
submanifold along which
a lot of
(maximally many) phases meet. This is a multiple ``point''.
Next we
compute the fine-structure constants
very close to this  ``point'', but
just barely in a convening phase that is consistent with
the survival of the Standard Model at long distances.
Do we get the correct fine-structure constants?

We did not really use Monte Carlo methods nor a group of 37
generators, but rather made analytic
estimates for what should have come out of
the Monte Carlo calculation outlined above.
These analytical calculations suggest that
there should be one or more
points in action parameter space where many phases meet. The MPP
asserts that gauge couplings have values
related to this(these) point(s).
In our approximation, the coordinates of this point are
found with the help of
the results of computer simulations done on some of the
subgroups of our Anti-GUT gauge group. We then calculate
what we would expect as the
gauge couplings extrapolated  down to long
distances.
The results that we obtain are in quite good agreement
with the experimental gauge coupling constants.

For use in calculating
the gauge couplings, we ideally want to determine
a point (or points) in the
action parameter space where many
phases meet for the whole Anti-GUT gauge group.
For the non-abelian subgroups \cite{db3}
SU(3) and SU(2), we found
that a good approximation to such a point
is obtained by
finding a multiple
point for each of the non-abelian invariant subgroups separately.
We then simply assume that the different
non-abelian groups do not interact at the
multiple point(s) for the whole
Anti-GUT gauge group.
This approach is straightforward
since there are
already published results with the
phase diagram of say an SU(3) gauge group
with both a triplet and an octet action term \cite{Dashen}.
There are three phases meeting at a point (the multiple point)
that  we interpret in
terms of what is the expected
behaviour of the fluctuations
very close to the Planck
or lattice scale.
These phases are
not really separate phases, in as far as
the one we interpret as the ``Coulomb phase''
is actually
connected to the ``confined'' phase by
going around the phase border
that ends at the tricritical point. The
third phase is one in which the ${\bf Z}_3$ subgroup of SU(3)
``confines'' while the continuous part of the group behaves in a
``Coulomb-like'' fashion.
The phase diagram is indeed very analogous
to the water-vapour-ice diagram,
where one also has the possibility of
going from vapour to water without
crossing a phase boundary by going
to sufficiently high temperature and pressure.
The interpretations we give in terms
of ``confinement'' and `` Coulomb'' are
only valid near the lattice scale,
while at long distances they are all really confining.
The phase that should be
identified phenomenologically with
``our phase'' is the one we denote as  ``Coulomb-like'',
because ``our phase'' should not be
confined in the high energy regime near the
Planck scale.

The SU(2) group has a quite analogous phase
diagram to SU(3), with three phases
meeting at a point in a phase diagram spanned by two action
parameters.

Now, as part of our scheme, the Anti-GUT gauge
group breaks down to the diagonal subgroup $SMG$.
The $SU(3)^3$ group breaks to its diagonal subgroup
$SU(3)_{diag}= \{(u,u,u) | u\in SU(3)\}$.
The inverse
fine-structure constant $1/\alpha_{3diag}$
for the diagonal subgroup is
given, in first approximation, as the sum
of the inverse fine-structure
constants for the three $SU(3)$ groups
in the Cartesian product subgroup $SU(3)^3$ of the
Anti-GUT gauge group.
The fine-structure constant of each of the three $SU(3)$ factors
is just that at the multiple point for
the single $SU(3)$ group, as if there
were no interaction between the
the three $SU(3)$'s (and other SMG$^3$ subgroups) that convene at the
multiple point.
It must be admitted that the formula for
the diagonal subgroup gauge coupling
\begin{equation}
1/\alpha_{3diag} = 1/\alpha_{3, 1st\;gen.} + 1/\alpha_{3,2nd\; gen.}
+ 1/\alpha_{3,3rd \; gen.}
\label{alphadiag}
\end{equation}
is not invariant in going from one
scheme of renormalisation to another
one, so it can hardly be very accurate.
We hope that if we use it for
schemes like MOM or $\overline{MS}$
or schemes numerically not far from them,
then it should hold approximately,
while we do not expect it to work in
a true lattice scheme.  So we either rewrite the
lattice scale couplings
corresponding to the multiple point for $SU(3)$
in one of the schemes in which
we trust this formula, or we have to really
figure out a more accurate formula.
In reality, we assumed that Parisi-correcting the lattice
couplings would bring us to one of the schemes
for which it should  be approximately trustable.
The formula is easily derived upon recalling that
the Parisi correction
\begin{equation}
1/g^2_{Parisi\; corrected} = \frac{<trU>}{Tr1}  \cdot 1/g^2
\end{equation}
supposedly gives you an effective action in a continuum formulation.

Then of course the couplings obtained by Parisi correction
must be the running couplings, referred roughly
to the lattice scale
which is only an order of magnitude or so above
the scale at which the Anti-GUT
group is Higgsed to the diagonal
subgroup, by the $W,S,T$ and $\xi$ fields
discussed above. Therefore a renormalisation group running
down to the experimentally
accessible scale is needed.
Only in this stage of the calculation
do we take into account the influence of fermions;
we have assumed it a good approximation
that this is the only influence of the fermions
on the phase border couplings.

For the U(1)'s it is suspected that we do not find the highest number
of meeting phases simply by
treating the different U(1) groups independently.
Rather we invented \cite{db1}
a coupling between the three
$U(1)$ subgroups of SMG$^3$ (the
extra $U(1)_f$ has been ignored in this
calculation) that has a discrete symmetry of the same structure
as a certain hexagonal lattice in
a three dimensional parameter space. The lattice
with this symmetry has nothing to do
with the space-time lattice in the
lattice regularization. It is rather a lattice in the covering group
${\bf R }^3$ of the group $U(1)^3$, which is
the abelian part of our Anti-GUT
gauge group once the $U(1)_f$
is ignored. Having this discrete symmetry,
we can use it to permute the
various anticipated phases into each other.
By imposing the symmetry
just at the hoped for multiple point,
we can be sure that if a phase
is in contact with the multiple point,
then  the images of this phase under the symmetry are
guaranteed to meet there also.
In this way we ensure that a rather large number of phases
can convene at the multiple point, provided we
can construct an action that both has this symmetry
and, at the same time, can bring together at the multiple point
just some of the hoped for phases. But this requires the
inclusion of extra action terms and, thereby, introduces the risk
that these are required to have such large coefficients
that they cause other unwanted
phases to dominate.

The reason that interactions are
presumably more relevant for the
abelian than for the non-abelian
groups is that there are more invariant
subgroups for a Cartesian product of
abelian groups than for a Cartesian product
of simple groups. For abelian groups, all
subgroups are automatically invariant.
The relevance of invariant subgroups is
that they can confine by themselves leaving
the rest (i.e. the factor group) in
a  ``Coulomb-like'' phase. It is also
easier to have interactions between the various
abelian subgroup factors of a Cartesian
product than in the non-abelian case. This is
because the interactions one could have
for the non-abelian subgroups would
require irrelevant terms that would really be
regularization (lattice) artifacts.
Such interactions are not easy to
make in the continuum. For the abelian groups however,
one can even have interactions between different U(1) groups
as relevant terms:
\begin{equation}
const. \cdot F_{\mu\nu}^{1st \; gen.}(x) F^{\mu\nu \; 2nd\; gen.}(x).
\end{equation}
With such terms in the Lagrangian density, we
might arrange that some linear
combination of first and second generation
$U(1)$ fields could make confining
fluctuations while another combination
behaves Coulomb-like. By using such
possibilities there is the chance
of constructing an interaction such that
a large number of phases can be made to meet.
But how many and which phases actually
meet might be sensitive to our
approximations.

As we want the Lagrangian for the special case in which all
three abelian fields are equal---corresponding
to the diagonal subgroup to be identified
with the weak hypercharge group in the Standard
Model---such
interaction terms will contribute to the final $1/\alpha_{1diag}$.
In the picture that we believe corresponds
to the `best' multiple point
it happens, mainly due to the
hexagonal symmetry, that the coefficients
of the interaction terms are just like the one on the
usual $ {(F_{\mu\nu}^{1st\; gen.})}^2$ term.
These terms add up so that we get
six terms (3 squares and three interaction terms) leading to the
$1/\alpha_{1diag}$ being 6 times
as big as the $1/\alpha_1 $ found at the
multiple point in the diagram for a
single $U(1)$ (for the non-abelian
case there was instead only a factor 3).
Actually, including the corrections from
the extra interaction terms leads to a factor between 6 and 7.

Hence we obtained the following values for the
fine-structure constants:
\begin{equation}
\begin{array}{lll}
   & \mbox{predicted} & \mbox{experimental} \\
\alpha^{-1}_3(M_Z) & 12\pm 6 & 9.25\pm 0.43 \\
\alpha^{-1}_2(M_Z) & 29\pm 6 & 30.10\pm 0.23 \\
\alpha^{-1}_1(M_Z) & 99\pm 5 & 98.70\pm 0.23 \\
\alpha^{-1}(0) & 137\pm 9 & 137.036...
\\
\label{alphas}
\end{array}
\end{equation}
It should be stressed that the good agreement
seen in eq.~(\ref{alphas}) is
achieved using a pure desert renormalisation
group extrapolation of the Parisi corrected etc.\cite{db1}
Planck scale predictions.
The theoretical uncertainties, typically of the
order $\pm 6 $, for the inverse fine-structure
constants are crude estimates
of the reliability of our going to the continuum
and also include an estimate of the uncertainty on the
Monte Carlo data used.
If one introduced supersymmetry broken around
the weak scale, the modification of the running
of the fine-structure
constants would cause our predictions for the
inverse fine-structure constants---at the $Z^0$ scale---to
go up by about 15 to 20. This corresponds
to about 3 standard deviations
{\em for each of the three inverse
fine-structure constants}. That is to say,
since we have already (accidentally)
very good agreement, inclusion
of supersymmetry down to the
weak scale would weaken our coupling
constant predictions by about 3
standard deviations for each of the three
couplings!

So our model, although not by itself necessarily
in conflict with supersymmetry
at ``low'' energies, would no longer fit the data with such supersymmetry.
Therefore we predict that
supersymmetric
partners should {\em not} be found in any experiments that are even
remotely realistic
in our time.

As can be seen from figure 1, our
predictions are at the Planck scale
and thus can only be tested together
with an extrapolation model, which
we take to be the desert (minimal Standard Model) almost all the way
up to the Planck scale. Since the running is
also sensitive to the number of
generations, our predicted low energy
couplings depend on this number, both
because of the renormalisation group
running and because of the number
of $SMG$'s in the Anti-GUT gauge group
being  equal to the number of
generations $N_{gen}$. In terms of $N_{gen}$,
our gauge group is written
$SMG^{N_{gen}}\times U(1)^k$ (where the
number $k$ of $U(1)_f$ groups
varies in a more complicated way,
but is k=1 for $N_{gen}=2$ and 3, and  $k=2$ for
$N_{gen}$= 4). It turns out that both
dependences make our predicted couplings
at the electroweak scale weaker the larger $N_{gen}$ is.
It is, therefore, not surprising that even an
older, less sophisticated version \cite{picek} of
the fine-structure constant calculation part of our model led
to a fit indicating that the number of generations
$N_{gen}$ must be three in order that our couplings can fit.
In this work, only the non-abelian
couplings are treated even approximately the
way we do in the present model.
Our approach was to see if our fit of the number
of generations $N_{gen}$ led to an integer, namely 3.
It did so and, in this sense,
a version of our model showed that
it could truly {\em pre}dict the number
of generations, at a time when
only cosmological fits indicated that
3 was a little better than 4 in
fitting the abundances of the
primordially produced isotopes.

\section{Conclusion; \quad Are There Any
Chances of Realistic Tests?}

We have put forward a model, or scheme,
for physics at the Planck scale
having a gauge group
$SMG^3\times U(1)_f$, as we call it, and in
which it is imposed that the
coupling constants be so as to make
several vacua degenerate in energy density.
In addition we made
very data-inspired choices for the Higgs
fields that break down this
37-dimensional group to the Standard Model group, in
the sense of choosing their
quantum numbers under the big 37-dimensional group. We used four
such Higgs fields, one of which $S$ has a VEV
of order unity in Planck units, and
the Weinberg-Salam Higgs field $\phi_{WS}$ which
also must be assigned quantum numbers under
the bigger group.

The fundamental Yukawa couplings in our model are taken
to be of order of magnitude unity. We then obtained a fit
to the quark-lepton masses and mixing matrix, in terms of
the Higgs field VEVs. One of the VEVs was fixed of order unity,
$<S> = 1$, in Planck units. So, in addition to the
electroweak scale $<\phi_{WS}> = 246$ GeV, we
used just three new parameters---namely the $W$, $T$ and $\xi$
Higgs field VEV suppression factors. We have
no explanation for the order of magnitude of $<\phi_{WS}>$: the
gauge hierarchy problem remains the most difficult to solve.
Otherwise we fit, order of magnitudewise or better, the 19
parameters of the Standard Model. If we take this
rather impressive fit as a signal for truth, it is in a way
slightly sad in the sense that our model has, in its uncorrupted
form, no new physics except the Weinberg-Salam Higgs particle
until the Planck scale. So there are not
many positive predictions, rather the negative one
that you shall find no new physics at accessible
scales: No SUSY at experimental scales, no right handed $SU(2)$
etc. rather only the dull Standard Model.

One may wonder: are our predictions
really so dull that there is almost no
way to settle, by further investigation, whether
our model should be right or wrong?
Well, there might be a few chances:

$\bullet$    Cosmological Strings that can split.

The $U(1)^4$ subgroup in our model is broken down to
the single $U(1)$ of the Standard Model. So the Higgs fields
in one point of space-time must formally
give rise to a non-simply connected
space of configurations compatible with being a vacuum locally.
This local configuration space has a $\Pi_1$
homotopy group ${\bf Z}^3$ which should
give rise to several types of stable vortex
lines or cosmological strings, and
the different types of string would be able to branch into
each other. So there would be a network of
such cosmological strings, rather
than just a single unbranched type of string as is usually considered.
This feature may have some cosmological consequences which
could be looked for.

$\bullet$ Baryogenesis problem.

In our model we have, essentially at least, just
the Standard Model interactions
up to the Planck scale or, rather, one
or two orders of magnitude below it.
So we have no way, at the electroweak scale, to generate the
phenomenologically determined
number of about $10^{-9}$ baryon per photon
in the cosmic background radiation. There
is insufficient CP violation in the Standard Model.
Furthermore, even if created, it would immediately
be washed out by sphaleron transitions after the electroweak
phase transition since we already know from LEP that
the Standard Model Higgs boson mass is greater than 70 GeV.
So the only chance in our scheme to get a sufficient number of
baryons relative to the photon number is to postulate that there is,
at some stage, a violation of the
quantum number $B-L$ (= baryon number minus lepton number).
This quantum number $B-L$ is anomaly free and
exactly conserved by the Standard Model.
However, there is
an anomaly for this quantum number in our model
due to the $U(1)_f$ gauge field. That is to say
that, at temperature scales so high that
the $U(1)_f$ subgroup is unbroken,
there is $B-L$ violation due to the anomaly.
So, barring CP-violation and
deviation from equilibrium, there will be a wash-out
of the $B-L$ to zero even if there
were some truly primordial $B-L$.
The calculation of any surviving $B-L$ and, consequently, the
baryon number produced in our model (after $B+L$ is
washed-out by electroweak sphaleron transitions)
is a future challenge for us. Really
it looks crudely as if we reach
a number about four orders of magnitude
too low (but we should calculate
more carefully):

According to our scenario there is, in the beginning,
an anomaly in $B-L$ conservation mainly due
to the $U(1)_f$ gauge field. This anomaly
keeps washing out any net $B-L$ that
might appear, due to CP-violating forces from
the Planck scale physics, until the temperature
of the Universe has fallen so low that
the $\xi$ and other Higgs fields get
their VEVs.
The non-Standard Model gauge particles of the $SMG^3\times U(1)_f$
group then obtain masses and the $B-L$ quantum number
becomes much better conserved.  However $B-L$ is still
violated by irrelevant terms, reflecting
the Planck scale physics where
``everything'' happens. We imagine that there are
effective irrelevant terms, breaking $B-L$ as well as CP for example,
active at the temperature when the physics has
already come to first approximation into the desert.
A term in the Lagrangian of dimension $d=5$
which violates $B-L$ conservation is
\begin{equation}
g\bar{l^c} l \phi\phi + h.c.
\end{equation}
where $l$ is a left-handed lepton field, e.g. the
($\tau$, $\nu_{\tau}$) field,
and $\phi$ is the Weinberg-Salam Higgs field.
By using a coupling $g$ which
is complex, CP can be violated in this term.
Applying this term in lowest order perturbation theory,
there is still no CP-violation in the rates of the
processes governed by it. However by interference
with a next order term,
it is possible to obtain a difference in the cross-sections of say
$\phi^+ \tau^- \rightarrow \phi^- \tau^+$
and the CP conjugate process
$\phi^- \tau^+ \rightarrow \phi^+ \tau^-$.

Although with time-reversal breaking there
does not have to be detailed balance,
the numbers of particles of different
sorts will of course be (statistically) stable in
thermal equilibrium. In an expanding Universe, however, the
density falls and, since the scattering typically
delays the out-going
particles a bit, the rates in various processes may no
longer keep the various species in chemical equilibrium.
We should imagine that even in thermal equilibrium, but with
time-reversal broken, there is a stationary circulation of particles
or pairs of particles around say three channels.
For example the particles $\phi^+ \tau^-$
may scatter more often into $\phi^- \tau^+$
than the opposite way
$\phi^- \tau^+ \rightarrow \phi^+ \tau^-$.
But then there is an opposite reaction going on
{\em indirectly} via some third channel. Such a third channel may be
achieved by using several flavours.
For example the particles could circulate through
the chain of processes:
\begin{equation}
\phi^+ \tau^- \rightarrow \phi^- \mu^+ \rightarrow
\phi^- e^+ \rightarrow \phi^+ \tau^-
\end{equation}
Small time delays
\footnote{If the energy dependence of a scattering amplitude is Fourier
transformed into a function of what is a
passage time variable, one easily gets a delay
in the sense of some extra averaged
passage time needed for the process}
cause some of the
processes in the circulation chain to last
a little longer than the others and, in
the presence of the Hubble expansion, a non-equilibrium
results, which can even lead to an excess of $B-L$.

In our estimate of the $B-L$ number produced, we shall assume
the cross sections for the processes in this ``circulation''
were of order unity in Planck units. So the rate for, say,
the process $\phi^+ \tau^- \rightarrow \phi^- \tau^+$
would be given by the probability of finding
two particles to scatter within a Planck
length and that is $T^6$, where T is
the temperature (in Planck units).
For the moment we ignore the suppressions due to the
selection rules in  our model, except for the need to
use the $d=5$ terms. Then we estimate the time-reversal
violation to be lower than the process rate by
yet another factor of T.
Ignoring the number of degrees of freedom factor, the time
in the early Universe is given by $t = 1/T^2$.
Hence the rate $T^7$ for the B-L asymmetric (under CP)
violation gives a net production of $T^5_{inital}$, where
$T_{initial} $ is the temperature at the moment when the
$B-L$ anomaly switches off and production can start.

However, it is not at all correct,
in our model, that the irrelevant term processes here used
are of order unity in Planck units even if, as we assume,
``everything'' except gauge symmetry violation goes on at the
Planck scale.
These processes are in fact suppressed
by the approximate conservation of the global charges
corresponding to our $SMG^3\times U(1)_f$ gauge group,
even after the Higgs fields
W, T, $\xi$ and $S$ have got their non-zero VEVs.
The rates are therefore suppressed
similarly to the fermion
masses and mixing angles, by such factors as W, T
\footnote{It should be clear from the context whether $T$
refers to a Higgs field suppression factor or
temperature.} etc. It happens that
the terms we need are rather closely related to the neutrino mass
matrix elements. A very preliminary estimate suggests
a suppression factor $W^2 T$ for the
$\phi^+ \tau^- \rightarrow \phi^- \mu^+$
transition, in analogy to the mass term in the neutrino
mass matrix connecting the $\mu$ neutrino with the
$\tau$ antineutrino.
The next process
$\phi^- \mu^+ \rightarrow \phi^- e^+$
in the ``circulation'' described above
has a suppression factor of $\xi^3$
in our model.
The final step $\phi^- e^+ \rightarrow \phi^+ \tau^-$
then needed acquires a suppression factor $W^2T\xi^3$,
analogous to an off-diagonal
(Majorana) neutrino mass matrix element.
So putting $T_{initial} = \xi$ we get the $B-L$ density
to be
\begin{displaymath}
\xi^5 \times ``suppression factors''= \xi^{11} W^4T^2
\approx 0.099^{11} \times 0.179^4 \times  0.071^2 = 10^{-16.3}
\end{displaymath}
at the time corresponding to this temperature.
Now conventionally this is measured relative
to the entropy density,
also roughly the density of photons, which
at this time is $T_{initial}^3 \approx 10^{-3}$.
Thus the baryon number to photon number ratio
is in our model predicted to be $10^{-13.3}$,
which is not so terribly bad compared to the phenomenological
value of $ 10^{-9}$.
Perhaps one may even find a more copious
mode of production within
our model.

$\bullet$  Octet companions of the Weinberg-Salam Higgs.

The Weinberg-Salam Higgs must
lie in an irreducible
representation of the full gauge group,
but this representation is not necessarily
also irreducible under
the Standard Model Gauge group. So we can
expect partners or companions of
the Weinberg-Salam Higgs field.
It is possible the mysterious reason for the Weinberg-Salam
Higgs components getting such a small mass scale
may also extend to the
other components of the same $SMG^3\times U(1)_f$
irreducible representation. In this case there could be
practically observable companions of the Weinberg-Salam
Higgs field. It happens that in our detailed model
the Weinberg-Salam Higgs representation
is indeed reducible under the Standard Model
and a colour octet companion
is predicted. Actually it is
an SU(2) doublet and there will thus
be both a neutral and
a charged colour octet companion.
They should be looked for experimentally.
Since they are octets under QCD-colour they cause no
baryon non-conservation, unlike the analogous
triplet companion in SU(5)-GUT. Thus the octet
companions could be very light, without making
themselves felt through proton decay.

$\bullet$ Discrete group flux strings.

One of the major principles in our model---MPP---is that
several phases of the vacuum should coexist.
Now, in some phases, discrete subgroups confine
while the continuous part of the group does not
follow suit.
A priori the flux tubes representing the discrete gauge fields
have string tensions of order unity in Planck units.
However at the phase boundary it just passes through zero
and it could be that the phase transition was so ``weak'' that
this tension was exceptionally small.
In that case there could be experimentally detectable
effects on, say, the $ Z \rightarrow b \bar{b}$ vertex
and the string states could be sufficiently light
to be detected.

$\bullet$  Neutrino oscillations may give access to
very short distance physics.

Since neutrino oscillations can reflect an exceedingly high mass
see-saw particle,
we have via neutrino
oscillations a window to the physics at very high energies.
So it may touch on the validity of our model
in a way that does not just ``see''
the desert and its pure Standard Model interactions.
In the unmodified model, we have a desert
up to the Planck scale apart from the VEVs of our
Higgs fields W, T and $\xi$, i.e.~to just one
or at most two orders of
magnitude under the Planck scale.
We have also made more detailed neutrino
Majorana mass estimates in our model.
But in the most clean version \cite{smg3m}
with the see-saw mass scale set by the Planck mass,
we predict the neutrinos to be too light to give any practically
observable neutrino oscillations!
Taken this way our model is already
{\em falsified } by the present neutrino oscillation experiments.
We have, however, proposed how to modify
our model with just one extra Higgs field.
In this very lightly
modified version, we predict the solar neutrinos to come in just
{\em half} the amount predicted by the no oscillation
calculations \cite{neutrinoself}.

$\bullet$ Several vacua may be found; non-local effects.

If really it is so important that many
vacua have the same energy density,
you should expect
these many vacua to be realized somewhere or sometime.
That is to say we  might look for some new
type of vacuum spreading through the
Universe as a bubble. There might be a
chance to see some galaxies or the
like being mirrored in the surface of
such a bubble. Presumably it would
move with a speed near that of light and
you would see the mirrored galaxies
blue shifted. Or should we make a bubble ourselves?

The vacua could come to exist first
in the future but, anyway, it seems
hard to get a general model making
these degenerate vacua unless you use
either

1) non-local effects at the fundamental level,
like in baby universe theory.
In fact we can argue that just having non-locality in the mild form,
that the coupling constants are influenced by an average of the fields
over both future and past, easily leads to the degenerate vacuum
principle. So non-local effects might be what are predicted to get
our phenomenological principle explained.
\newline Or

2) Supersymmetry, which easily gives
different vacua with different physics
but the same (zero) energy density.

It should be stressed though that our
detailed predictions would be
spoiled by supersymmetry at experimental scales at least.
So a supersymmetry explanation of
the degenerate vacua would have to
be based on a very strongly broken
supersymmetry---presumably broken at the Planck scale.

\section*{}
\footnoterule
{\em \footnotesize
To be published in the Proceedings of the APCTP-ICTP Joint
International Conference (AIJIC 97) on Recent Developments in
Nonperturbative Quantum Field Theory, Seoul, Korea,
26-30 May 1997.
}
\end{document}